\documentclass[aps,prd,showpacs,twocolumn,10pt,superscriptaddress,floatfix,nofootinbib,longbibliography]{revtex4-1}

\usepackage{amssymb,amsfonts,amsmath,bm,bbm}
\usepackage{graphicx}
\usepackage{dcolumn}
\usepackage{bm}
\usepackage{CJK} 
\usepackage{subfigure}  
\usepackage{float} 
\usepackage[
colorlinks=true,
linkcolor=blue,
breaklinks=true,
urlcolor=blue,  
citecolor=blue]{hyperref}
\usepackage{bbold}
\usepackage{epstopdf}

\newcommand{\HNUST}{\affiliation{
Hunan Provincial Key Laboratory of Intelligent Sensors and Advanced Sensor Materials, \\ School of Physics and Electronics,
Hunan University of Science and Technology, \\Xiangtan  
 411201, China}}

\newcommand{\SXU}{\affiliation{Department of Physics and Institute of Theoretical Physics, Shanxi University, \\Taiyuan 030006, China}}

\newcommand{\UCAS}{\affiliation{
School of Nuclear Science and Technology,
      University of Chinese Academy of Sciences,
      \\Beijing 100049, China}}

\newcommand{\IHEP}{\affiliation{
Institute of High Energy Physics, Chinese Academy of Sciences, Beijing 100049, China}}

\newcommand{\LZU}{\affiliation{
School of Nuclear Science and Technology,
      Lanzhou University, Lanzhou 730000, China}}

\newcommand{\ANU}{\affiliation{
School of Physics and Electrical Engineering,
      Anyang Normal University, Anyang 455000, China}}

\begin{document}

\title{Landau quantization and spin polarization
of cold magnetized quark matter}

\author{Zhen-Yan Lu
}
\email{luzhenyan@hnust.edu.cn}\HNUST

\author{Jian-Feng Xu
} \email{jfxu@aynu.edu.cn}\ANU

\author{Xin-Jian~Wen
}\email{wenxj@sxu.edu.cn}\SXU

\author{Guang-Xiong Peng
} \email{gxpeng@ucas.ac.cn}\UCAS\IHEP

\author{Marco Ruggieri}
\email{ruggieri@lzu.edu.cn}\LZU

\date{\today}

\begin{abstract}
The magnetic field and density behaviors of various thermodynamic quantities of strange quark matter under compact star conditions are investigated in the framework of the thermodynamically self-consistent quasiparticle model.
For individual species, a larger number density $n_i$ leads to a larger  magnetic field strength threshold
that align all particles
parallel or antiparallel to the magnetic field.
Accordingly,
in contrast to the finite baryon density effect which reduces the spin polarization of magnetized strange quark matter, the magnetic field effect leads to an enhancement of it. We also compute the sound velocity as a function of the baryon density and find the sound velocity shows an obvious oscillation with increasing density. Except for the oscillation, similar to the zero-magnetic field case that the sound velocity grows with increasing density and approaches the conformal limit $V_s^2=1/3$ at high densities from below.
\end{abstract}

\maketitle

\section{Introduction} \label{INTRO}

Quarks are generally believed to appear within
hadronic matter, due to the feature
of color confinement.
However, it is expected, with increasing density, that the basic constituent of hadronic matter, i.e. hadrons,
might squeeze out to form deconfined quark matter at low temperatures.
About forty years ago, Witten first conjectured that strange quark matter (SQM)~\cite{Farhi-1984qu}, consisting of roughly equal number of
up, down, and strange quarks,
could be the true ground state of strong interaction~\cite{Witten-1984rs}.

Since SQM could be absolutely stable, there might be compact objects with a quark core or even
completely made of quarks and leptons, i.e., the so-called strange stars~\cite{Haensel-1986qb,Olinto-1986je,Alcock-1986hz,Xu-2000sr,Weber-2004kj}.
Recently,
comparison of the theoretical calculations in a model-independent way with astrophysical observations suggests that
there might be a quark matter core in the maximally massive neutron stars~\cite{Annala-2019puf}.
Several studies also imply that strange stars can coexist with neutron stars~\cite{Bombaci-2000cv,Drago-2015cea,Bhattacharyya-2017mdh,Drago-2013fsa,Drago-2015dea,Bombaci-2020vgw,Drago-2020gqn,Lourenco-2021lpn}.
One important feature of such stars is the associated strong magnetic field,
with strength of
the order $10^{11}\sim10^{13}$
G on the surface of pulsars~\cite{Haensel-2007yy}, or even reaches $10^{14}\sim 10^{15}$ G in some magnetars~\cite{Duncan-1992hi}.
In the core
of compact stars, the
magnetic
field strength is estimated as high as $10^{18}\sim 10^{20}$ G~\cite{91Lai-APJ,Bocquet-1995je,Broderick-2001qw,Ferrer-2010wz,Ferrer-2012wa}.
Except for the
stellar objects, the magnetic field strength produced in the heavy ion collisions can
be as strong as $10^{18}\sim 10^{19}$ G~\cite{08Kharzeev.McLerran.ea227-253NPA,Fukushima-2008xe,Skokov-2009qp,Voronyuk-2011jd,Deng-2012pc}.
However, let us emphasize that estimates of the largest value of magnetic field $B_m$ inside the compact stars exist, see for example Ref.~\cite{Chatterjee-2014qsa} and references therein; however, these estimates have some dependence on the model used for the equation of state of bulk matter.
In this study we take the $B_m=O(10^{18}~\mathrm{G})$ as a conservative estimate of the maximum value of $B_m$,
and use larger values of $B_m$ only for illustrative purposes, see Fig.~\ref{fig:BmUpDown2n04n0}.

The presence of a strong magnetic field introduces Landau levels for the charged particles
and hence has significant effects on the stability of SQM~\cite{Chakrabarty-1996te,Isayev-2015jqa,Menezes-2009uc}.
The properties of SQM in a strong magnetic field have attracted increasing interest over the past few decades,
for example, the effects of an external magnetic field on the symmetry energy~\cite{Chu-2018dch,Chu-2020afg},
equation of state~\cite{Isayev-2018hzq} and surface tension of quark matter~\cite{Lugones-2018qgu}, and $r$-mode instability~\cite{Huang-2009ue,Xu-2021eyi} as well as structure properties of strange stars~\cite{Hou-2014pba,PeresMenezes-2015ukv,Chu-2014foa,Kayanikhoo-2019ugo,Chu-2021oey}, etc.
In addition, the presence of a magnetic
field can also affect the in-medium
chiral condensates~\cite{Menezes-2008qt}, and hence the critical temperature of the chiral phase transition~\cite{Ruggieri-2016xww,Wen-2021mgm,Ferrer-2020tlz,Xu-2020yag,Chaudhuri-2019lbw,Chaudhuri-2020lga,Grunfeld-2014qfa,Farias-2021fci}, as well as the deconfinement
transition~\cite{Fraga-2012fs,Mizher-2010zb,Ferreira-2013tba,11Gatto.Ruggieri34016-34016PRD,Backes-2021mdt}.
Due to the intractable nature of quantum chromodynamics (QCD) in the nonperturbative regime,
we need to find a proper way to mimic the strong interactions between quarks.
The medium-dependence of quarks
can be
taken into account by considering a chemical potential and/or temperature dependent quark mass in the quasiparticle model~\cite{Peshier-1994zf}. In this case, however, special attention should be paid to the thermodynamic consistency
of the phenomenological models
~\cite{Gorenstein-1995vm,Bannur-2006js,Alba-2014lda,Xu-2014zea,Xu-2015wya,Xia-2014zaa}.
To confine quarks, an effective bag constant is required in the nonperturbative regime\footnote{Recently, there was a discussion on the quark mean-field model for nuclear matter with or without a bag constant and
quark confinement is found to be mainly demonstrated by the bag after it is included in the model, instead of the confining potential~\cite{Zhu-2018vwn}.}. Since the quark masses are chemical potential and/or temperature dependent, the additional effective bag constant should also be simultaneously chemical potential and/or temperature dependent in order to satisfy the fundamental relation of thermodynamics~\cite{Peshier-1999ww,Ma-2018bwf}.
Following the original idea in Refs.~\cite{Schertler-1996tq,Schertler-1997za}, the quasiparticle model with a fixed strong coupling has been used to study the finite-size strangelets~\cite{Wen-2009zza} and the medium effects on the surface tension of strangelets~\cite{Wen-2010zz,Wen-2011zz}.
Later, this model is
extended to include the running strong coupling with a chemical potential dependent renormalization subtraction point, which is constrained by the Cauchy condition in the chemical potential space~\cite{Lu-2016fki}.

In this work, we restrict ourselves to the deconfined  SQM and study the effect of a magnetic field on the properties of SQM, especially
on the behavior of
maximum Landau levels, particle  fractions, relative spin polarization, and sound velocity
at finite
baryon densities
in the presence of a uniform external magnetic field.
We can anticipate the main new results of our study: within the quasiparticle model, we study for the first time the polarization of dense strange quark matter in a strong magnetic background field, and we complete this investigation by the calculation of the squared speed of sound.
The paper is outlined as follows. In Sec.~\ref{secQPM}, we review the thermodynamic consistent quasiparticle model with a chemical potential dependent bag constant in strong magnetic fields.
In Sec.~\ref{secDiscussion}, we  present the numerical results and give discussions
for our calculations.
Finally, a summary is given in Sec.~\ref{CONCLUSION}.

\section{Thermodynamic consistent quasiparticle model} \label{secQPM}

Unlike the density-dependent mass model
that the quark masses are density-dependent~\cite{Chakrabarty-1991ui,Peng-1999gh,Wen-2005uf,Lu-2016jsv,Chu-2012rd},
in the quasiparticle model, the strong interactions between quarks
are mimicked by arranging a chemical potential dependence of quark masses.
For the medium dependence, the effective quark mass used in this work was derived in the zero momentum limit of the dispersion relations following from an effective quark propagator~\cite{Schertler-1996tq} by resuming one-loop self-energy diagrams in the hard dense loop approximation~\cite{Vija-1994is,Weldon-1982aq}
\begin{equation}\label{eq:QuakmS}
m_{i}^{*}
=\frac{m_{i 0}}{2}+\sqrt{\frac{m_{i 0}^{2}}{4}+\frac{g^{2} \mu_{i}^{2}}{6 \pi^{2}}},
\end{equation}
where $m_{i0}$ and $\mu_i$ are respectively the current quark mass and chemical potential of quark flavor $i$.

In Refs.~\cite{Wen-2009zza,Zhang-2021qhl}, the properties of strange quark matter with and without finite size effects are investigated in the quasiparticle model, in which the strong coupling is treated as a pure constant.
However, it is well-known that the strong coupling $g$ is
running with the energy scale,
and here we adopt the following phenomenological expression~\cite{Patra-1995gs}
\begin{equation}\label{eqALPHAs}
g^{2}=
\frac{48 \pi^{2}}{29\ln \left(a \mu_{i}^{2}/\Lambda^{2}\right)},
\end{equation}
where $a=0.8$, and $\Lambda$ is
the QCD scale parameter
controlling the rate at which QCD coupling runs as a function of energy scale.
In the vanishing current mass limit,
Eq.~(\ref{eq:QuakmS}) reduces to
\begin{equation}\label{eq:mUD}
m_{i}^{*}=\frac{g \mu_{i}}{\sqrt{6} \pi},
\end{equation}
which can be used as the effective masses for up and down quarks since their current masses are small compared to the strange quark. Note that since electrons do not participate in the strong interactions, their masses keep constant as in the normal case.

To consider the effects of a strong magnetic field on the Landau quantization and the spin polarization of the strongly interacting quark matter, we need to know the energy spectrum of the spin-half particle in a strong magnetic field. Without loss of generality, we assume the external magnetic field along with the $z$-direction, i.e., $\boldsymbol{\mathcal{B}}=B_m\hat{z}$.
In consequence, the energy spectrum of the spin-half charged particle can be obtained by solving the Dirac equation. We accordingly have
\begin{eqnarray}\label{eq:Efexpression}
\varepsilon_{i}=\sqrt{p_{z}^{2}+\bar{m}_{i, \nu}^{*2}
}.
\end{eqnarray}
Here $\bar{m}_{i, \nu}^{*}=\sqrt{m_{i}^{*2}+2 \nu |q_i| B_m}$ is the effective mass of particle $i$ in the presence of an external magnetic field, $p_z$ is the particle momentum in the $z$-direction,
$q_i$ is
the electronic charge of particle $i$,
and the Landau level $\nu$ is defined as~\cite{Broderick-2000pe,Strickland-2012vu}
\begin{eqnarray}\label{eq:nu}
\nu=l+\frac{1}{2}-\frac{\eta}{2}\frac{q_i}{|q_i|},
\end{eqnarray}
where $l$ denotes the orbital angular momentum, and $\eta=\pm 1$ represents the two eigenstates of the spin-half charged particle with ``+1" for spin up and ``-1" for spin down.
Due to the Landau quantization,
the integral over the momentum components perpendicular to the magnetic field become discrete. Consequently, we have
\begin{equation}\label{eq:INTsum}
\int
\int
\int
d p_{x} d p_{y}d p_{z} \rightarrow 2 \pi |q_i| B_{m} \sum_{\eta=\pm 1} \sum_{l}\int \mathrm{d} p_{z}.
\end{equation}
Rewrite Eq.~(\ref{eq:Efexpression}) in terms of the chemical potential, and define the maximum $p_z$ as $p_{z,F}=\sqrt{\mu_{i}^{2}-\bar{m}_{i, \nu}^{*2}}$, we then have  an upper limit for the Landau levels
\begin{equation}\label{eq:Nmax}
\nu \leq \nu_{i,\max } =
\text{Int}\left[\frac{\mu_i^{2}-m_i^{*2}}{2|q_i| B_m}\right]
\end{equation}
due to the fact that the Fermi momenta $p_{z,F}$ must be real-valued quantities.
In Eq.~(\ref{eq:Nmax}), the symbol $\text{Int}[...]$ represents 
floor of the enclosed quantity.


At zero temperature,
the quasiparticle contribution $\Omega_i\equiv\Omega_i(\mu_i,m_i^{*})$ to the total thermodynamic potential density for magnetized quark matter
is given by
\begin{eqnarray}\label{eq:Omegai}
\Omega_{i}
&=&
-\frac{d_{i} |q_i| B_m}{4 \pi^{2}} \sum_{\eta=\pm 1} \sum_{l}\Bigg\{ \mu_{i} \sqrt{\mu_{i}^{ 2}-\bar{m}_{i, \nu}^{*2}}\nonumber\\
&&-\bar{m}_{i, \nu}^{*2} \ln \Bigg(\frac{\mu_{i}+\sqrt{\mu_{i}^{ 2}-\bar{m}_{i, \nu}^{*2}}}{\bar{m}_{i, \nu}^{*}}\Bigg)\Bigg\},
\end{eqnarray}
with the degenerate factor $d_i=1$ for electron and $d_i=3$ for quarks respectively.
The pressure and energy density for magnetized SQM within the quasiparticle model
are given by
\begin{eqnarray}
P&=&-\Omega-B^{*}, \\
E&=&\Omega+\sum_{i} \mu_{i} n_{i}+B^{*}, \label{eq:EE}
\end{eqnarray}
where $\Omega\equiv\sum_i\Omega_i$ is the total thermodynamic potential density\footnote{A term $B_m^2/2$ coming from the magnetic field contribution has been dropped
since it is irrelevant for the present work.} containing a summation of all quasiparticle contributions in Eq.~(\ref{eq:Omegai}),
while
$B^{*}\equiv \sum_iB_i(\mu_i,m_i^{*})+B_0$
is the chemical potential dependence of effective bag constant introduced to fulfill the thermodynamic self-consistency requirement~\cite{Wen-2009zza}. $B_0$ is the MIT bag constant, which is taken to be zero since it is not numerically relevant to our calculations.
Note that there is no need to introduce an effective bag constant for electrons since they do not participate in the strong interactions and thus do not have a chemical potential dependent mass.
For a quasiparticle Fermi system the number density of the component $i$ has the same form of the free-particle case in the presence of an external magnetic field, which is given by
\begin{eqnarray}\label{eqnQ}
n_{i}=\frac{d_{i} |q_i| B_{m}}{2 \pi^{2}} \sum_{\eta=\pm 1} \sum_{l}
\sqrt{\mu_{i}^{2}-\bar{m}_{i, \nu}^{*2}
}.
\end{eqnarray}
According to the fundamental differential equation of thermodynamics, the number density $n_i$ is obtained by taking the first derivative of the thermodynamic potential density with respect to the corresponding chemical potential $\mu_i$.
Mathematically,
this is equivalent to requiring
\begin{eqnarray}\label{eq:QPM}
n_{i}=-\left.\frac{\mathrm{d} \Omega}{\mathrm{d} \mu_{i}}\right|_{\mu_{j\neq i}}
&=&-\frac{\partial \Omega_{i}}{\partial \mu_{i}}-
\underbrace{\left[\frac{\partial \Omega_{i}}{\partial m_{i}^{*}} \frac{\mathrm{d} m_{i}^{*}}{\mathrm{d} \mu_{i}}+\frac{\partial B^{*}}{\partial \mu_{i}} \right]}_{=0}.~~
\end{eqnarray}
In the above, the first term on the right hand side of the second equality is in fact equal to Eq.~(\ref{eqnQ}), while the vanishing of second term is required to satisfy the thermodynamic consistency requirement of the quasiparticle model~\cite{Lu-2016fki}.
From Eq.~(\ref{eq:QPM}), we have
\begin{eqnarray}
\frac{\mathrm{d} B^{*}}{\mathrm{d} \mu_{i}} \frac{\mathrm{d} \mu_{i}}{\mathrm{~d} m_{i}^{*}}=-\frac{\partial \Omega_{i}}{\partial m_{i}^{*}}
\end{eqnarray}
or equivalently
\begin{eqnarray}\label{eq:effBag}
B^{*}
&=&-\frac{d_{i} |q_{i}| B_{m}}{2 \pi^{2}} \sum_{\eta=\pm 1} \sum_{l} \int_{\mu_{i}^{c}}^{\mu_{i}}
m_{i}^{*} \frac{\mathrm{d} m_{i}^{*}
}{\mathrm{d} \mu_{i}}
\nonumber\\
&& \times \ln \Bigg(\frac{\mu_{i}+\sqrt{\mu_{i}^{2}-\bar{m}_{i, \nu}^{*2}}}{\bar{m}_{i, \nu}^{*}}\Bigg) \mathrm{d} \mu_{i}.
\end{eqnarray}
To ensure the positive of square root in Eq.~(\ref{eq:effBag}),
the lower limit of the integration over $\mu_i$ in Eq.~(\ref{eq:effBag}) should satisfy
\begin{equation}
\mu_{i}^{c 2}-\bar{m}_{i, \nu}^{*2} \geq 0.
\end{equation}
By using the effective quark mass in Eq.~(\ref{eq:QuakmS}), the medium dependent effective bag constant $B^{*}$ can be derived by numerical integration.

\section{Numerical Results and discussions}  \label{secDiscussion}

In this section, we will present our results on the properties of
various thermodynamic quantities
in the presence of an external strong magnetic field at nonzero baryon densities.
For the stable SQM, the $\beta$-equilibrium can be reached by the weak reactions $d, s \leftrightarrow u+e+\bar{v}_{e}$ and $\quad s+u \leftrightarrow u+d$.
Correspondingly, we have the following conditions
for the relevant chemical potentials as
\begin{eqnarray}\label{eq:chemical}
\begin{cases}
\mu_d=\mu_s\equiv \mu,  \cr
\mu_u+\mu_e=\mu.
\end{cases}
\end{eqnarray}
Here the chemical potential of neutrinos is set to zero because they can enter or leave the system freely.
We also have the expressions of the
baryon density
\begin{eqnarray}\label{eq:nb}
n_b=\frac{1}{3}(n_u+n_d+n_s),
\end{eqnarray}
and the charge neutrality condition
\begin{eqnarray}\label{eq:neutrility}
\frac{2}{3}n_u-\frac{1}{3}n_d-\frac{1}{3}n_s-n_e=0,
\end{eqnarray}
which has to be fulfilled for stable SQM presented in compact stars.

\subsection{Chemical potentials and Landau levels}\label{sec:chemicals}

\begin{figure}[bt]
\centering
  \includegraphics[width=247pt]{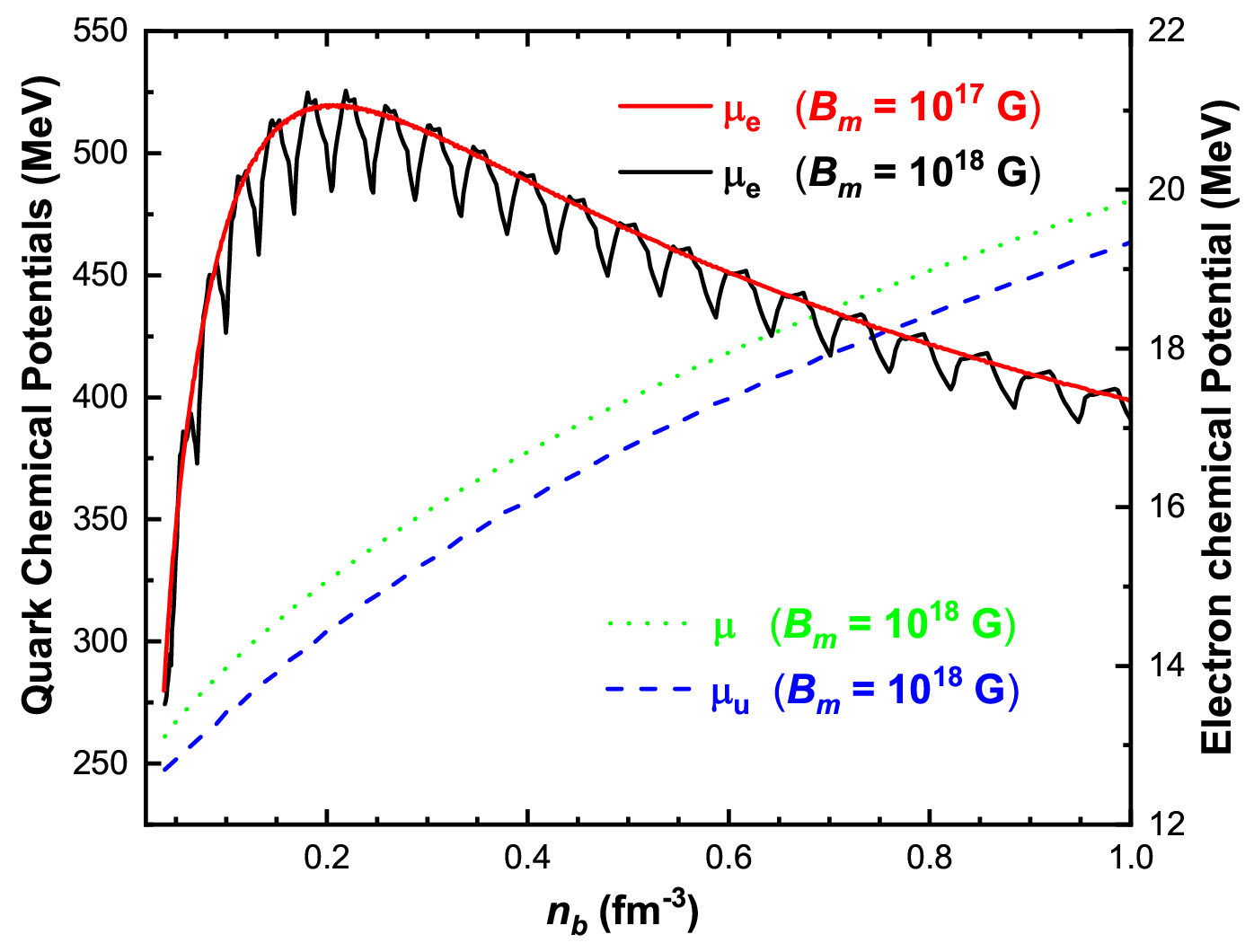} \\
  \caption{(Color online) Quark and electron chemical potentials as functions of the baryon density for two different magnetic field strengths.
 }\label{fig:nbChemical}
\end{figure}

\begin{figure*}
  \includegraphics[width=0.8\textwidth]{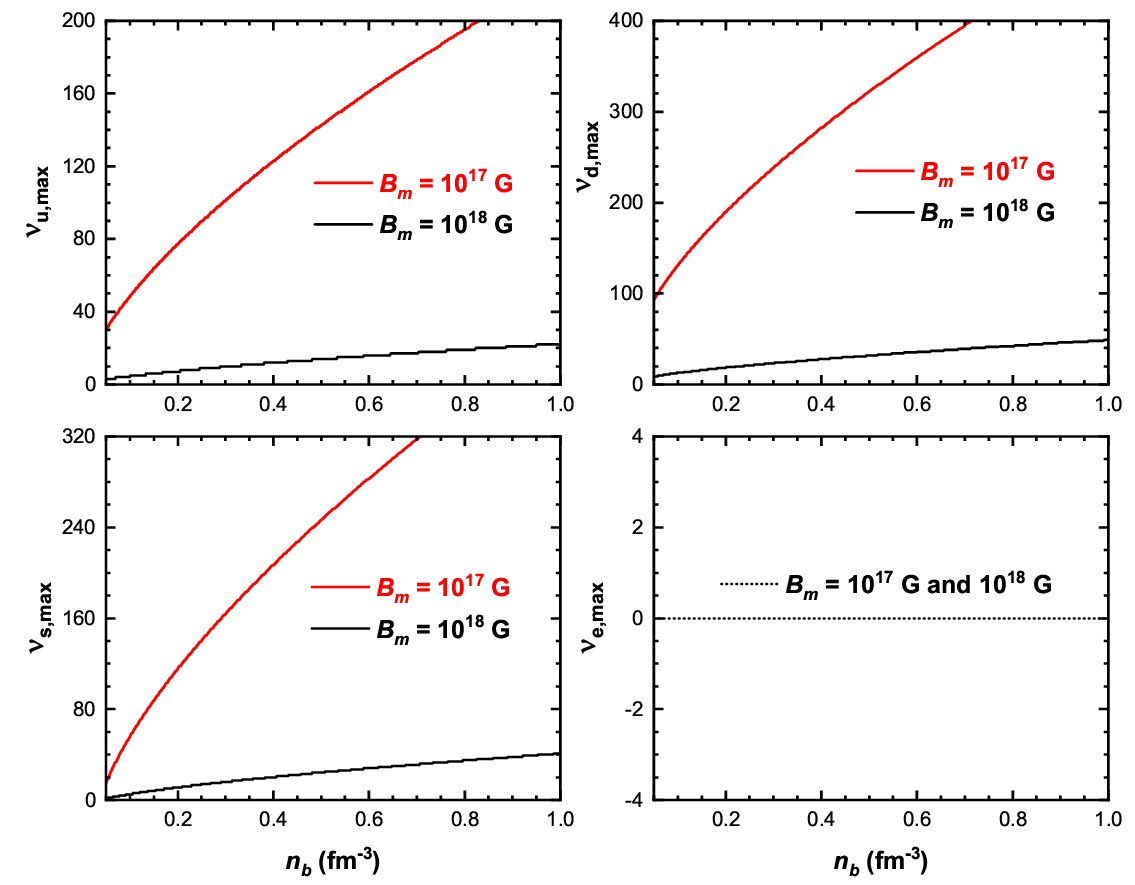}\\
  \caption{(Color online) The maximum Landau levels for up, down, strange quarks, and electron change with the baryon density.
 }\label{fig:NmaxAll}
\end{figure*}

Strange stars are hypothetical objects
consisting of stable SQM in $\beta$-equilibrium condition.
For a given baryon density, one can numerically solve the set of equations in (\ref{eq:chemical}), (\ref{eq:nb}), and (\ref{eq:neutrility}) to obtain the corresponding quark and electron chemical potentials.
In Fig.~\ref{fig:nbChemical},
quark and electron chemical potentials are shown as functions of the baryon density at fixed $B_m$.
The blue dashed and green dotted curves correspond to $\mu_u$
and $\mu$ with the same magnetic field strength $B_m=10^{18}$ G,
while the red and black solid curves represent the electron chemical potentials at two different magnetic field strengths $B_m=10^{17}$ G and $B_m=10^{18}$ G, respectively. As can be seen from the figure, both $\mu_u$ and $\mu$ monotonically increase with baryon density for fixed magnetic field strength.
While
in both cases
the electron chemical potential $\mu_e$,
first increases 
 then decrease after reaching a maximum with increasing baryon density.
And the peaks for the electron chemical potential are approximately located at $n_b\approx0.20~\text{fm}^{-3}$.
The only difference of these two curves is the obvious oscillation shown by the solid black line with $B_m=10^{18}$ G, an order larger than $B_m=10^{17}$ indicated with the solid red line.
As a minor comment, we notice that $\mu_e$ experiences
fluctuations for large values of $B_m$.
These are due to the fact that electron density is fixed by the condition of electrical neutrality:
at zero temperature and in the lowest Landau level approximation, $n_e \propto eB_m\mu_e$; thus,
$\mu_e$ is sensitive to the behavior of the density of the quarks and in particular to their oscillations.
However, the oscillations of $\mu_e$ as well as those of the chemical potentials of the quarks are
very tiny, see the scale on the right vertical axis in Fig.~\ref{fig:nbChemical}: they are of the order of a few MeV,
therefore they are easy to visualize for the $\mu_e$ which is of the order of 20 MeV,
but are invisible for the quarks since their chemical potentials are in the range $200\sim500$ MeV.

Except for the
chemical potentials, the presence of a magnetic field also modifies the distribution of Landau levels.
In Fig.~\ref{fig:NmaxAll}, the maximum Landau levels for each component of
 SQM, i.e. up, down, strange quarks and electrons,
are plotted as functions of the baryon density for two different magnetic field strengths.
The red and black solid curves represent the cases with magnetic field strength  $B_m=10^{17}$ G and $B_m=10^{18}$ G, respectively.
We observe that the curves, representing the maximum Landau levels of quarks, grow almost linearly with increasing density at fixed magnetic field strength. Owing to the strong suppression of $\nu_{i,\text{max}}$ in strong magnetic fields, the slope of the black curves ($B_m=10^{18}$ G) is much smaller than that of the red curves ($B_m=10^{17}$ G).
Moreover, at very low densities with fixed magnetic field strength, the inequalities $\nu_{d,\text{max}}>\nu_{u,\text{max}}>\nu_{s,\text{max}}$ hold, while at relatively high density, we have $\nu_{d,\text{max}}>\nu_{s,\text{max}}>\nu_{u,\text{max}}$.
This phenomenon can be understood because down and strange quarks share the same chemical potential $\mu$, but the latter always has a larger effective mass due to the non-vanishing current mass $m_{s0}$, which automatically leads to $\nu_{d,\text{max}}>\nu_{s,\text{max}}$. On the other hand, the condition $\mu_u<\mu\equiv\mu_s=\mu_d$ should be fulfilled for any density because of the consideration of $\beta$-equilibrium with the weak reactions $d, s \leftrightarrow u+e+\bar{v}_{e}$ and $s+u \leftrightarrow u+d$. As a consequence, $\nu_{d,\text{max}}>\nu_{u,\text{max}}>\nu_{s,\text{max}}$ holds at low densities owing to the large value of effective strange quark mass but small strange quark chemical potential in this region. However, as we increase the density, the numerator in Eq.~(\ref{eq:Nmax}) for the strange quark increases much faster than the one for the up quark, and at the same time $\nu_{u,\text{max}}$ is always reduced by a factor of two compared to expression of $\nu_{s,\text{max}}$ due to the ratio $|q_u/q_s|=2$. It can be checked that as the density increases,
the difference between these two maximal Landau levels grows and finally leads to $\nu_{u,\text{max}}<\nu_{s,\text{max}}$ at high densities.

\begin{figure*}
   \includegraphics[width=247pt]{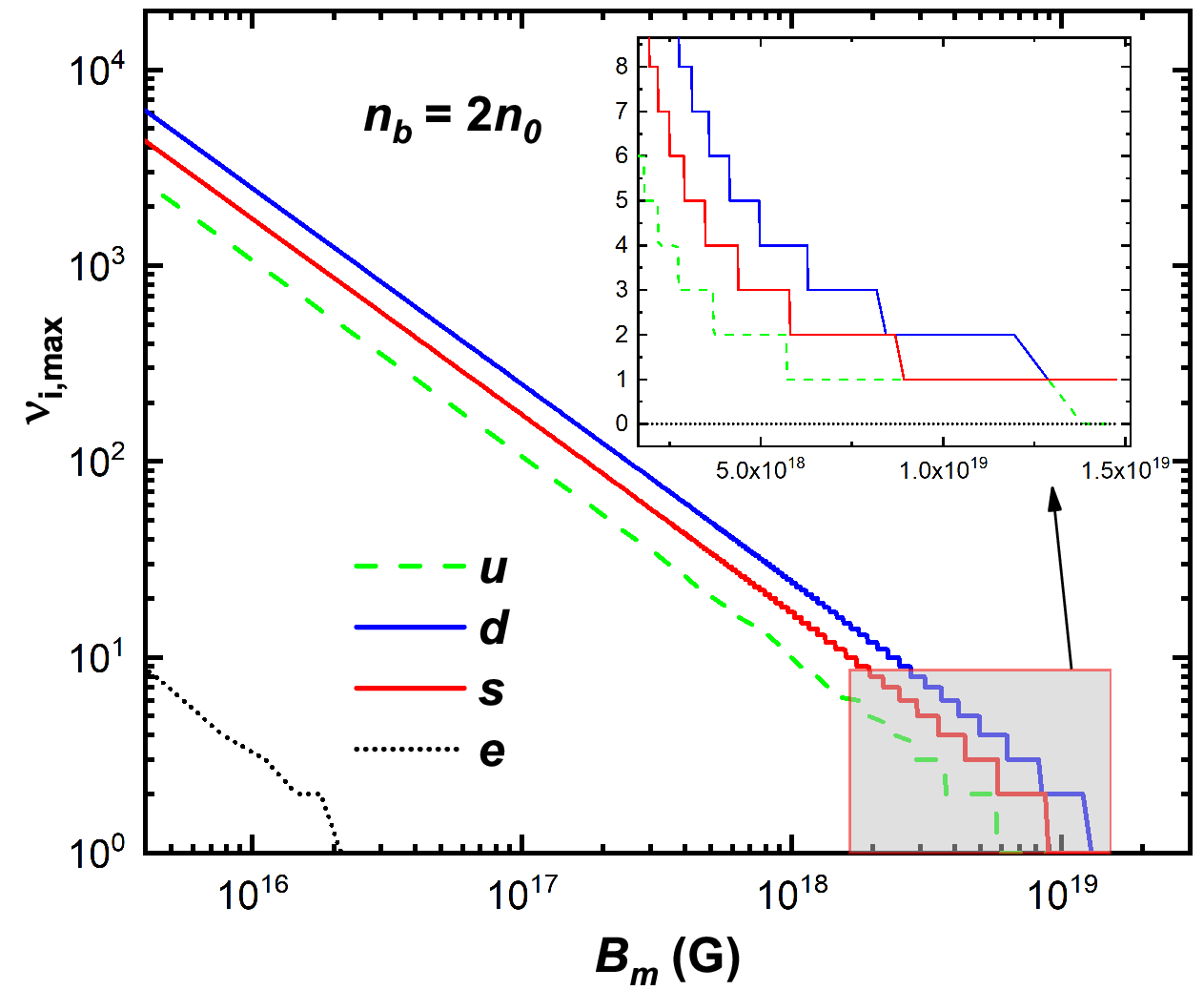}
  \includegraphics[width=247pt]{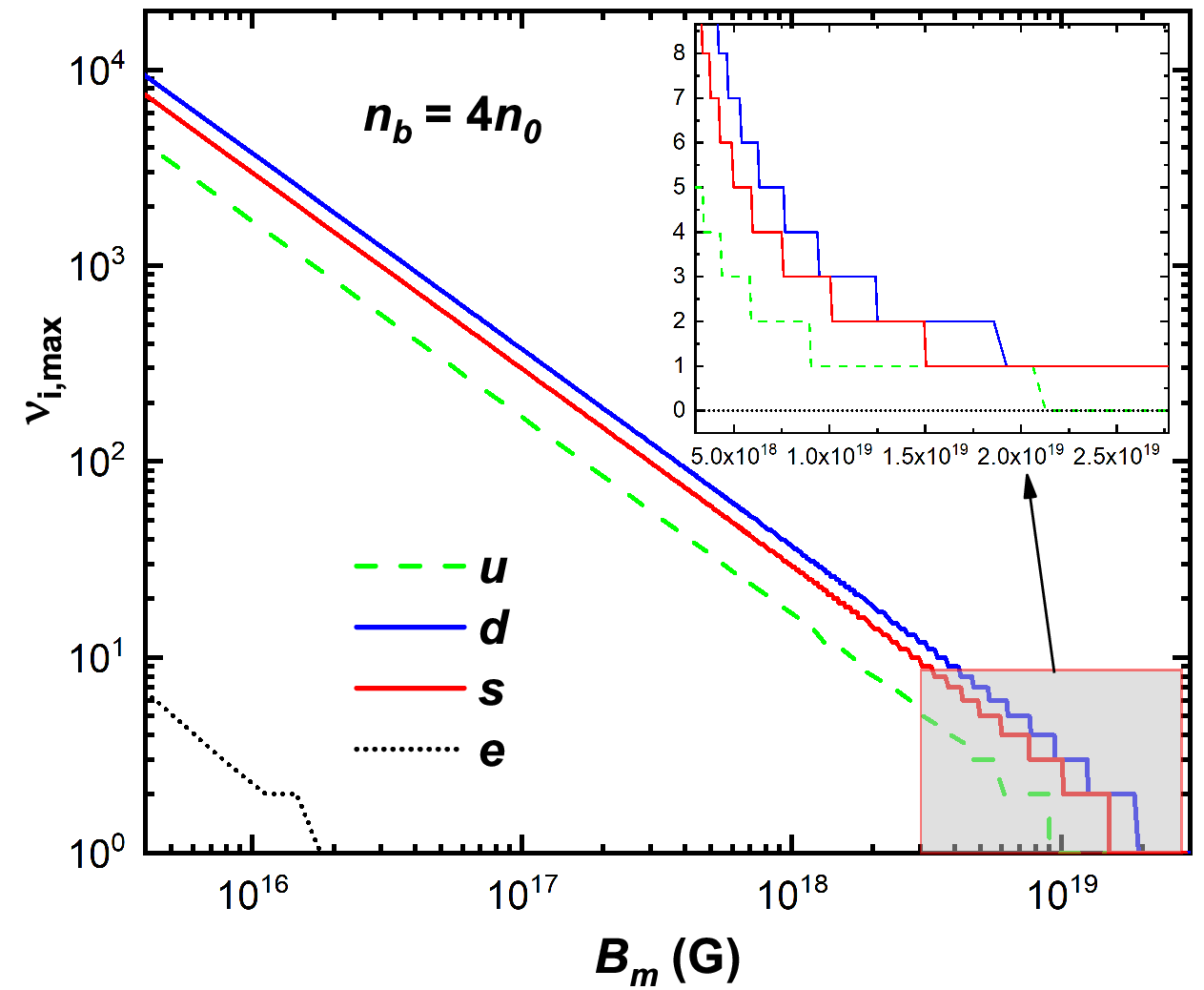}\\
  \caption{(Color online) The maximum Landau levels for various charged particles as a function of magnetic field strength for $n_b=2n_0$ (left panel) and $n_b=4n_0$ (right panel), where $n_0=0.16~\text{fm}^{-3}$ is the saturation nuclear density.
  }\label{fig:BmNmaxudse}
\end{figure*}

For the two selected magnetic field strengths, i.e., $B_m=10^{17}$ G and $B_m=10^{18}$ G, the maximum Landau levels for electrons keep zero since the corresponding electron chemical potentials are small, which are only a few tens MeV, compared with the magnetic field strengths. In this case,
only the lowest Landau level
contributes to the thermodynamic quantities of magnetized SQM, which is also explicitly confirmed in Fig.~\ref{fig:BmNmaxudse}.
This particular shape for the 
density dependence of the electron chemical potential is responsible for the behavior of the relative spin polarization $\Delta_e$ as shown in the two panels of Fig.~\ref{fig:BmUpDown2n04n0}.
From Eq.~(\ref{eqnQ}), it is expected that the number density of electrons
is completely determined by the behavior of the electron chemical potential $\mu_e$ for a fixed magnetic field strength, namely, $n_e$ also first increases and then decreases with increasing baryon density.

In Fig.~\ref{fig:BmNmaxudse}, we plot the maximum Landau levels for each component of $\beta$-equilibrium magnetized SQM as functions of the magnetic field strength
at $n_b=2n_0$ (left panel) and $n_b=4n_0$ (right panel), where  $n_0=0.16~\text{fm}^{-3}$ is the  saturation density of normal nuclear matter. For the sake of convenience we use the logarithm to label both the vertical and horizonal axes. 
The electron maximum Landau level $\nu_{e,\text{max}}$, represented by the black dotted curve, decreases from a value about ten to one
at around $1.8\times10^{16}~\text{G}$
and $1.5\times10^{16}~\text{G}$
in the left and right panels respectively.
While the maximum Landau levels for down, strange, and up quarks, which are denoted by the solid blue, solid red, and dashed green curves from right to left respectively, exhibit similar behavior: the curves
 decrease monotonously with increasing magnetic field strength and show obvious ladder-like shape at extremely large magnetic field strengths. Furthermore, as can be seen that the maximum Landau levels decrease almost linearly with the magnetic field strength, which agrees well with the Nambu$-$Jona-Lasinio finding of Ref.~\cite{Wen-2016atg}.

\begin{figure*}[bt]
   \includegraphics[width=247pt]{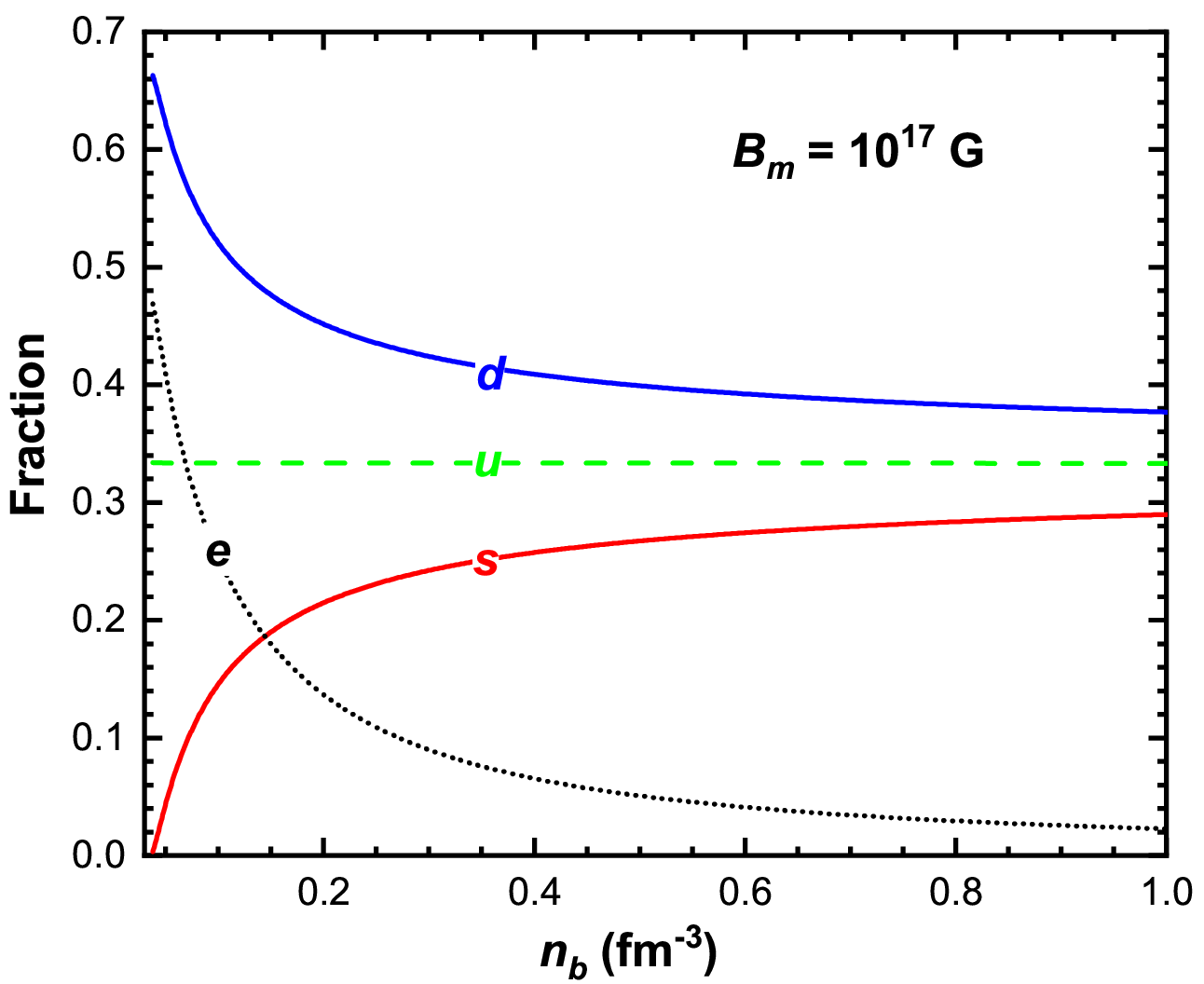}
  \includegraphics[width=247pt]{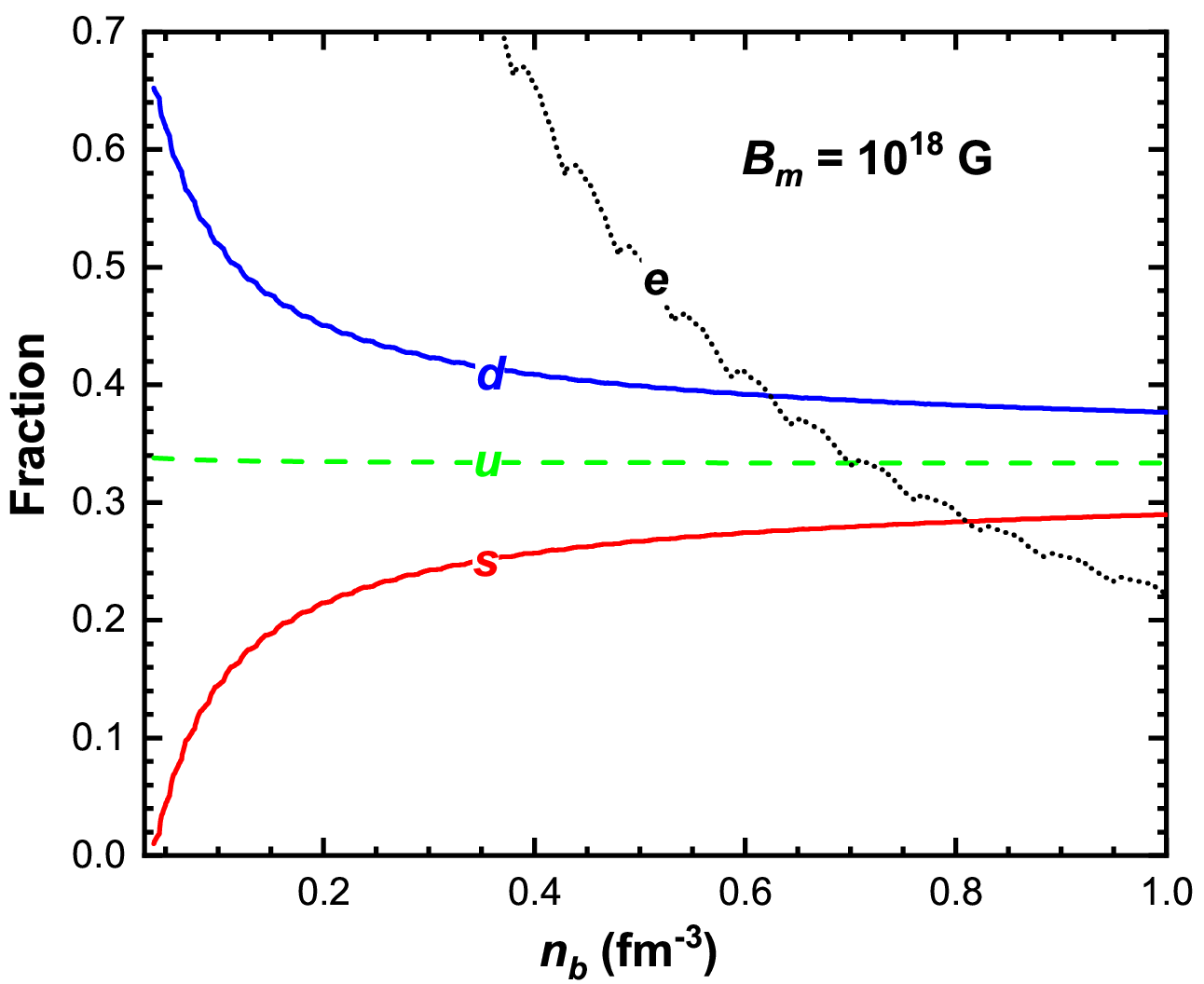}\\
  \caption{(Color online) Quark  fractions, $n_u/(3n_b)$, $n_d/(3n_b)$, $n_s/(3n_b)$ and $10^3$ times electron fraction $1000n_e/(3n_b)$,
   as functions of the baryon density for $B_m=10^{17}$ G (left panel) and $B_m=10^{18}$ G (right panel) respectively.
  }\label{fig:fractionsnB}
\end{figure*}

In Fig.~\ref{fig:fractionsnB}, we plot the quark fractions, i.e. $n_u/(3n_b)$, $n_d/(3n_b)$, $n_s/(3n_b)$, and the $10^3$ times electron fraction
$1000n_e/(3n_b)$, as functions of the baryon density at fixed magnetic field strength $B_m=10^{17}$ G (left panel) and $B_m=10^{18}$ G (right panel), respectively. One can see that the fraction of down and strange quarks decrease and increase with increasing baryon density respectively in both panels, while the fraction of up quarks almost keep constant in the considered range of the baryon density.
The fractions of different quark flavors approach each other when the density is large
enough. In addition, the fraction of electrons is very small and it decreases with increasing density. Also, the oscillation of the electron fraction become more obvious for larger magnetic field strengths.

\subsection{Spin polarization}

\begin{figure*}[bt]
\centering
  \includegraphics[width=247pt 
  ]{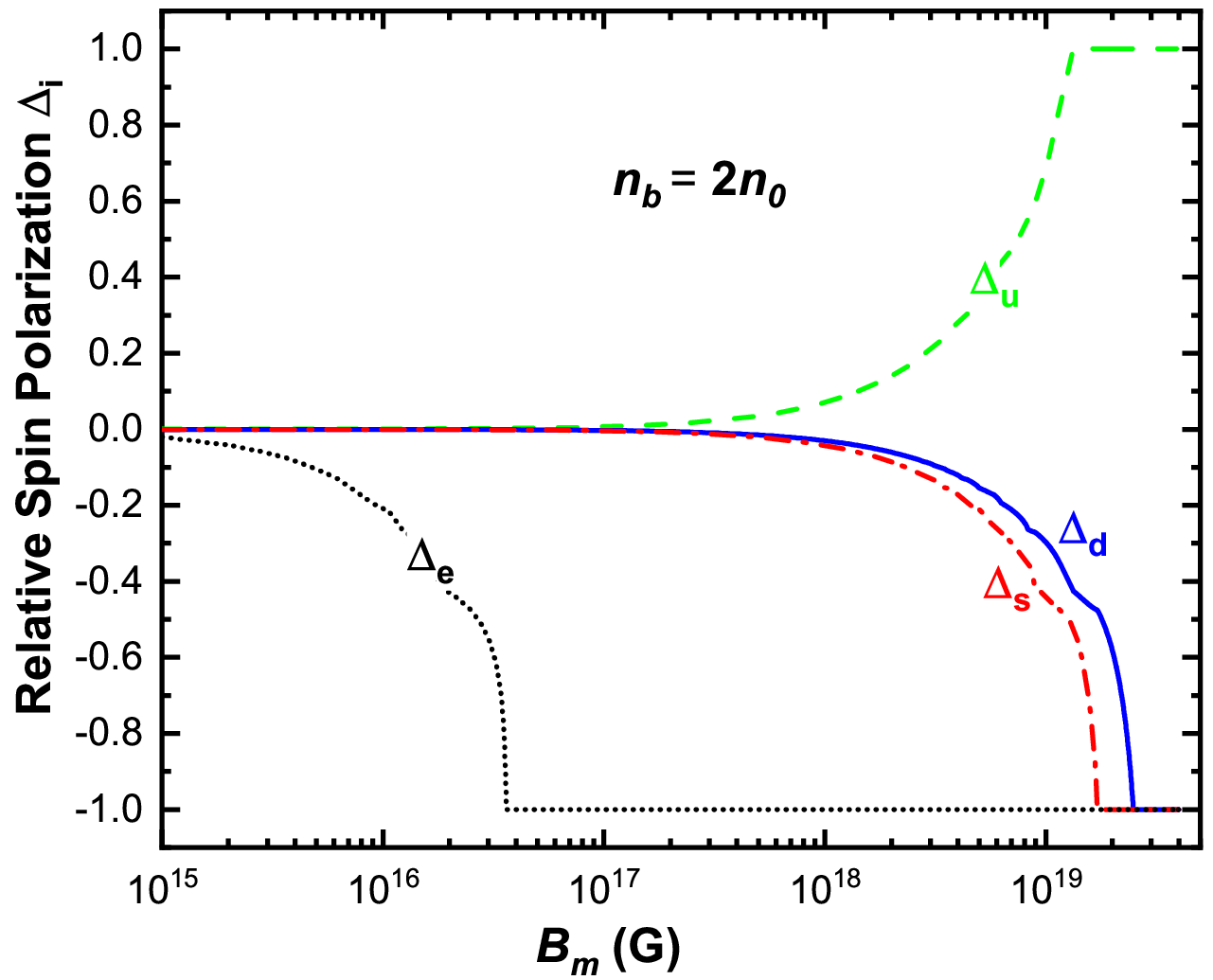}
  \includegraphics[width=247pt 
  ]{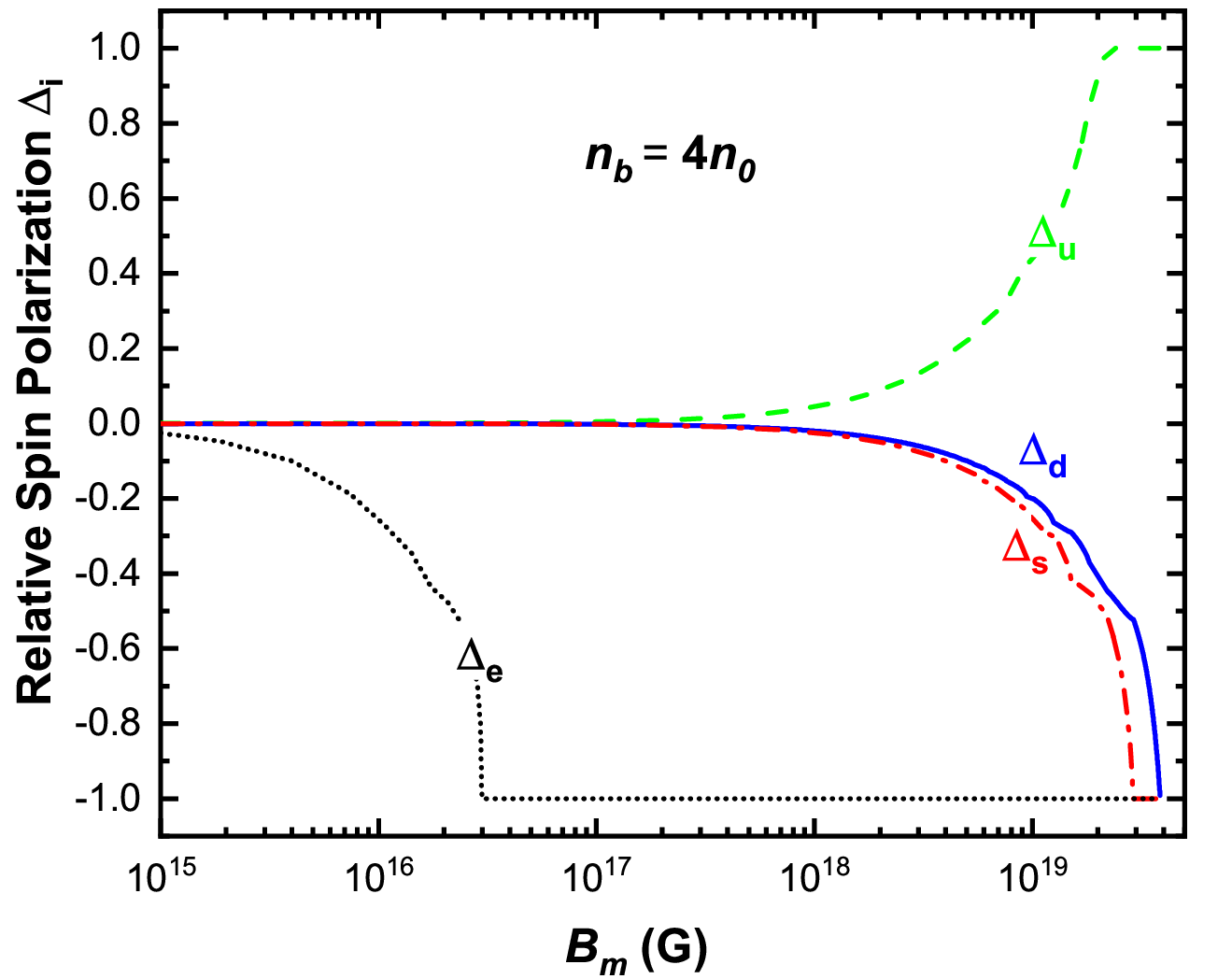}\\
  \caption{(Color online) Variation of the relative spin polarization for up, down, strange quarks, and electron 
  with respect to the magnetic field strength at $n_b=2n_0$ (left panel) and $n_b=4n_0$ (right panel).
 }\label{fig:BmUpDown2n04n0}
\end{figure*}

From Eq.~(\ref{eqnQ}), one can easily deduce that in the presence of an external magnetic field the number density of each constituent of magnetized SQM has two contributions, corresponding to the particles
with spin parallel and antiparallel to the magnetic field orientation respectively.
To study the spin polarization of the magnetized SQM,
we introduce the relative spin polarization
for $i$-type particle as~\cite{Perez-Garcia-2011czx,Avancini-2011zz,Rabhi-2014sza}
\begin{eqnarray}\label{eq:nudsUpDown}
\Delta_{i}=\frac{n_{i}^{\uparrow}-n_{i}^{\downarrow}}{n_{i}^{\uparrow}+n_{i}^{\downarrow}},
\end{eqnarray}
where $n_{i}^{\uparrow}$ and $n_{i}^{\downarrow}$ denote the number density of spin up and down $i$-type particles.
When $B_m=0$,
we obviously have $n_i^{\uparrow}=n_i^{\downarrow}$ and hence $\Delta_{i}=0$. 
For nonzero external magnetic field,
however, there is an upper value of the magnetic field $B_m^{c,i}$ for which complete
saturation of each constituent of the SQM occurs according to the expression $n_i=n_i^{\uparrow}+n_i^{\downarrow}$ and Eq.~(\ref{eqnQ}).
When the magnetic field reaches or excess the critical value $B_m^{c,i}$,
from Eq.~(\ref{eq:nu}) the condition for each constituent particle
$\Delta_{i}=1$ or $\Delta_{i}=-1 $
 is satisfied.
 The critical field value $B_m^{c,i}$ saturates the system and aligns all the $i$-type particles parallel or antiparallel to the magnetic field depending on the electric charge of the particles. Note that as deduced from the MIT bag model~\cite{Felipe-2007vb,Felipe-2008cm} without taking into account the strong interactions between quarks, the inclusion of anomalous magnetic moment will further affects the properties of magnetized SQM and hence alerts the values of $B_m^{c,i}$, but which is beyond the scope of this work.

In Fig.~\ref{fig:BmUpDown2n04n0}, we plot the relative spin polarization of each constituent of magnetized SQM as a function of the magnetic field strength, for two different baryon densities $n_b=2n_0$ (left panel) and $n_b=4n_0$ (right panel) respectively. The curves from top to bottom correspond to $\Delta_u$ (green dashed curve), $\Delta_d$ (blue solid curve), $\Delta_s$ (red dot-dashed curve), and $\Delta_e$ (black dotted curve) in both panels.
We observe that the curves in both panels show similar behaviors for each type particle with increasing magnetic field strength. More specifically, the relative spin polarization of electrons $\Delta_e$ begins to depart from zero at around $10^{15}$ G and the electrons become completely align antiparallel to the magnetic field at around $3.6\times10^{16}$ G and $3.0\times10^{16}$ G in the left and right panels, respectively.
While the relative spin polarization of quarks begins to departure from each other at around $3.0\times10^{17}$ G and this effect becomes notable after $10^{18}$ G. As shown in the left panel of Fig.~\ref{fig:BmUpDown2n04n0}, $\Delta_u$, $\Delta_s$, and $\Delta_d$ successively become completely saturated at the magnetic field strengths $1.3\times10^{19}$ G, $1.7\times10^{19}$ G, and $2.5\times10^{19}$ G.
And a comparison of these two panels shows that for a larger baryon density the critical magnetic field $B_m^{c,i}$
for quarks
becomes larger, whereas the one for electrons does
exactly the opposite.
As stated in Sec.~\ref{sec:chemicals}, this behavior
can be explained by the density dependence of the quark and electron chemical potentials 
shown in Fig.~\ref{fig:nbChemical}. Namely,
due to the monotonic behavior of quark chemical potentials as functions of the density, larger quark chemical potentials lead to
larger maximum Landau levels and in turn to smaller absolute values of $\Delta_{i}$ for quarks.
However, for the two chosen baryon densities $n_b=2n_0$ and $n_b=4n_0$, which are larger than typical value of the location of the peak for electron chemical potential $n_b\approx0.20~\text{fm}^{-3}$ shown in Fig.~\ref{fig:nbChemical}, the electron chemical potential decreases with increasing $n_b$, which means that a larger baryon density $n_b$ leads to a smaller maximum Landau level $\nu_{e,\text{max}}$ and hence to a smaller absolute value of $\Delta_e$.

Our study reveals that there is a narrow region, close to the critical field value that $\Delta_u=-1$ is reached, in which the $\beta$-equilibrium and charge neutrality conditions can not be fulfilled simultaneously. In other words, a negative solution for the electron chemical potential is found in this region~\cite{Isayev-2012sv}. When the magnetic field becomes larger, the charge neutrality condition is fulfilled again.
Besides, one can also read from the figure that depending on the electric charge, the up quarks tend to polarize
aligning spins with the magnetic field orientation, whereas the particles with negative electric charges like electrons, down and strange quarks do the opposite, which is consistent with the finding in Refs.~\cite{Perez-Garcia-2011czx,Avancini-2011zz}.

\begin{figure}[bt]
  \includegraphics[width=249pt]{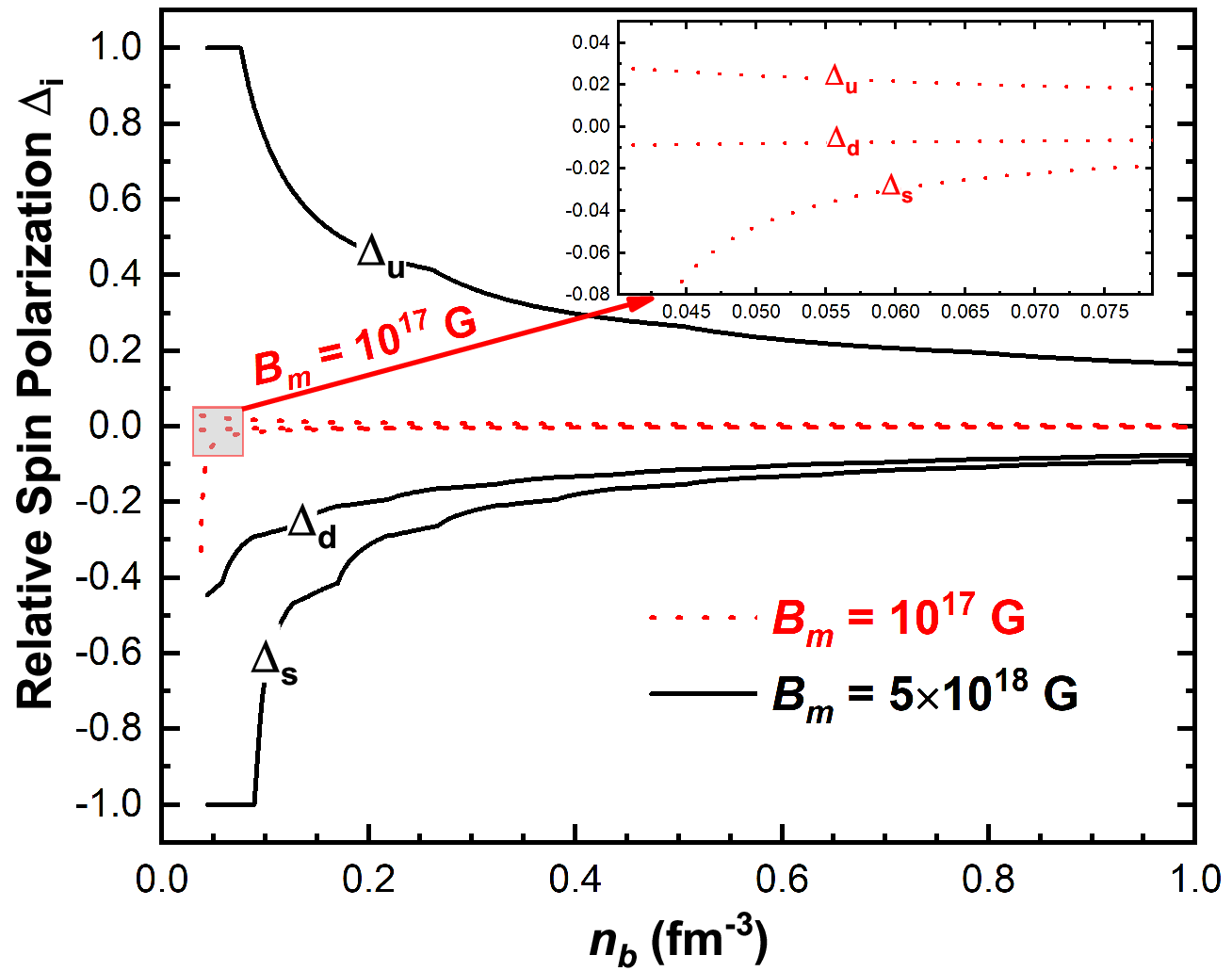}\\
  \caption{(Color online) Relative spin polarization of up, down, and strange quarks in $\beta$-equilibrium as a function of the baryon density for $B_m=10^{17}$ G (red dotted curves) and $B_m=5\times10^{18}$ G (black solid curves) respectively.
  }\label{fig:SpinPolarizationnB}
\end{figure}

In Fig.~\ref{fig:SpinPolarizationnB}, we display the variation of the relative spin polarization with density for up, down, and strange quarks
under the $\beta$-equilibrium condition at two representative magnetic field strengths $B_m=10^{17}$ G and $B_m=5\times10^{18}$ G.
For the magnetic field strength $B_m=10^{17}$ G,
denoted by the red dotted curves, the number densities with spin up and down
almost coincide with each other in the whole considered density range. While for the magnetic field strength $B_m=5\times10^{18}$ G,
denoted by the black solid curves, all the curves largely deviate from zero at low densities and gradually goes to zero with increasing baryon density. As we have shown in Figs.~\ref{fig:NmaxAll} and \ref{fig:BmNmaxudse} that the maximum Landau levels of quarks increase with the increase of the baryon density  but decrease with increasing magnetic field strength.
As a consequence,
the characteristic behavior of the occupation of the Landau levels responsible for
that although the magnetic field effects enhance the spin polarization, the finite baryon density effect tends to reduce the difference of spin polarization for magnetized SQM.

\subsection{Sound velocity}

\begin{figure}[bt]
  \includegraphics[width=249pt]{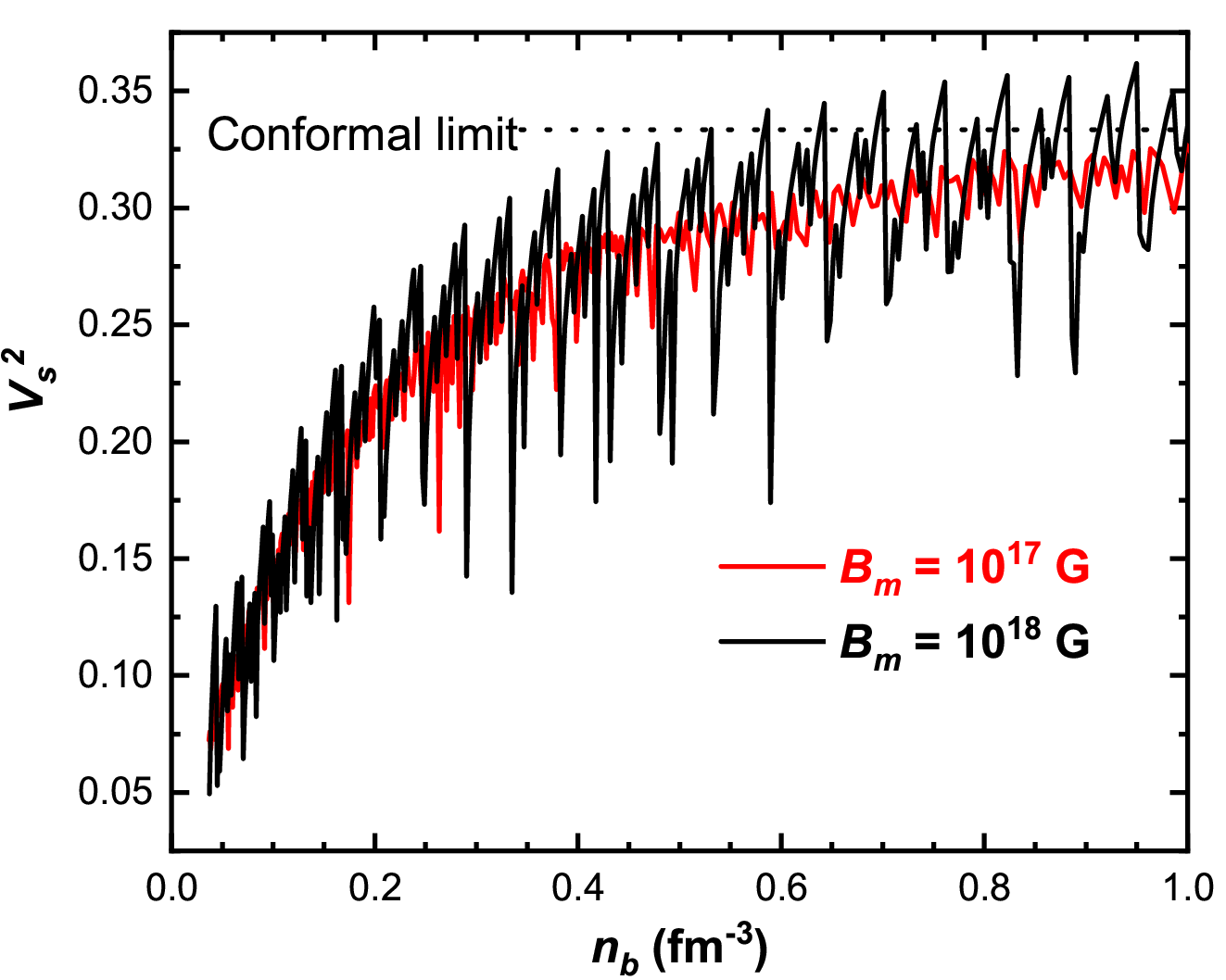}\\
  \caption{(Color online) Sound velocity of magnetized cold quark matter as a function of the baryon density for different values of magnetic fields. The horizontal line represents the conformal limit $V_s^2=1/3$.
  }\label{fig:soundVs}
\end{figure}

The strength of
interaction and the stiffness of the equation of state can be measured by the sound velocity.
In Fig.~\ref{fig:soundVs}, we present the sound velocity squared $V_s^2$ as a function of the baryon density, which is determined by
\begin{eqnarray}
V_s^2=
\frac{\mathrm{d} P}{\mathrm{d} E}.
\end{eqnarray}
The red and black solid curves correspond to $B_m=10^{17}$ G and $B_m=10^{18}$ G respectively, while the black dashed line represents the conformal limit.
It is found that in both cases, the sound velocity increase from small values with increasing baryon density and asymptotically approaches the conformal limit $V_s^2=1/3$ at high densities.
The only difference is that the curve with a stronger magnetic field shows a more obvious oscillation. Except for the oscillation, the density behavior of sound velocity in the present work is qualitatively the same as the zero magnetic field case~\cite{Wen-2009zza}. Since the two curves approaches the conformal limit from below, the sound velocity of magnetized SQM also fulfills the causality limit $V_s^2<1$.
It would be worth mentioning that
to explain the recent available cosmological data like large mass of the pulsars with a mass of about two solar masses
and tidal deformability observational data, there have been
many discussions suggesting that the sound velocity of dense matter should larger than the conformal limit $V_s^2=1/3$, see e.g. Refs.~\cite{Alsing-2017bbc,Tews-2018iwm,Reed-2019ezm}.  %
Such statements have been confirmed both by the model calculations~\cite{Xia-2019xax} and statistical analysis~\cite{Li-2021crp}.
However, we should emphasize that such kind of calculations are mostly done by assuming that the compact stars are neutron stars or hybrid stars
undergoing a phase transition~\cite{Hippert-2021gfs}.

\section{Conclusions}  \label{CONCLUSION}

We have explored the behaviors of various thermodynamic quantities of the magnetized SQM, with the consideration of $\beta$-equilibrium and charge neutrality conditions,
to the baryon density as well as to the magnetic field strength at nonzero chemical potentials
within the framework of the quasiparticle model.
The quark chemical potentials are found to increase
with the increment of baryon density at a certain magnetic field strength, which has the same behavior as the zero magnetic field case.

As the magnetic field increases, the relative spin polarization of electrons
and quarks begin to depart from zero at $\mathcal{O}(10^{15}~\text{G})$ and $\mathcal{O}(10^{17}~\text{G})$, and the corresponding absolute values become one at $\mathcal{O}(10^{16}~\text{G})$ and $\mathcal{O}(10^{19}~\text{G})$, respectively. In particular, for higher values of the baryon density the critical magnetic field strengths for $|\Delta_{i=u,d,s}|=1$ moves to larger values of $B_m$,
while the one for electrons does the opposite or the same depending on the location of the corresponding electron chemical potential as shown in Fig.~\ref{fig:nbChemical}. Namely, a smaller value of $\mu_e$ or $n_e$ leads to a larger $|\Delta_e|$ and in turn to a smaller critical magnetic field
when $\Delta_e=-1$ is satisfied.
Increasing the baryon density or the magnetic field affects the net polarization of quarks differently: this is easy to understand in terms of the occupation of the Landau levels, see Eq.~(\ref{eq:Nmax}). In fact, from that equation we see that increasing the chemical potential while keeping $B_m$ fixed results in a higher number of occupied Landau levels, thus reducing the polarization. On the other hand, increasing $B_m$ while keeping the chemical potential fixed, results in the lowering of the number of occupied Landau levels hence in the increase of the net polarization of strange quark matter.
Thus, one can conclude that
as a consequence of the occupation of Landau levels that
in contrast to the finite baryon density effect which reduces the absolute value of relative spin polarization,
the magnetic field effect leads to an enhancement of spin polarization of the magnetized SQM.

We have analyzed, for the first time within the quasiparticle model, the polarization of strange quark matter in the background of a strong magnetic field; moreover, we have also computed the squared speed of sound.
Compared to the zero-magnetic field case, the sound velocity in an external magnetic field shows an obvious oscillation with increasing density, especially for the curve with a stronger magnetic field. The sound velocity, in addition to the visible
oscillation phenomena, grows with increasing density and approaches the conformal limit $V_s^2=1/3$ at high densities from below.


\clearpage

\section*{Acknowledgments}

The authors thank the support from the
National Natural Science Foundation of China (Grant Nos.~11875181, 11875052,
11947098, 12005005, and 61973109),
the Hunan Provincial Natural Science Foundation of China (Grant No.~2021JJ40188),
the Scientific Research Fund of Hunan Provincial Education Department of China (Grant No.~19C0772), and the Scientific Research Fund of Hunan University of Science and Technology (Grant No.~E52059).
M. R. is supported by the National Natural Science Foundation of China (Grants No.~11805087 and No.~11875153) and by the Fundamental Research Funds for the Central Universities (Grant No.~862946).

\bibliographystyle{aapmrev4-2}
\bibliography{RefLuInsp}

\begin{thebibliography}{97}%
\makeatletter
\providecommand \@ifxundefined [1]{%
 \@ifx{#1\undefined}
}%
\providecommand \@ifnum [1]{%
 \ifnum #1\expandafter \@firstoftwo
 \else \expandafter \@secondoftwo
 \fi
}%
\providecommand \@ifx [1]{%
 \ifx #1\expandafter \@firstoftwo
 \else \expandafter \@secondoftwo
 \fi
}%
\providecommand \natexlab [1]{#1}%
\providecommand \enquote  [1]{``#1''}%
\providecommand \bibnamefont  [1]{#1}%
\providecommand \bibfnamefont [1]{#1}%
\providecommand \citenamefont [1]{#1}%
\providecommand \href@noop [0]{\@secondoftwo}%
\providecommand \href [0]{\begingroup \@sanitize@url \@href}%
\providecommand \@href[1]{\@@startlink{#1}\@@href}%
\providecommand \@@href[1]{\endgroup#1\@@endlink}%
\providecommand \@sanitize@url [0]{\catcode `\\12\catcode `\$12\catcode
  `\&12\catcode `\#12\catcode `\^12\catcode `\_12\catcode `\%12\relax}%
\providecommand \@@startlink[1]{}%
\providecommand \@@endlink[0]{}%
\providecommand \url  [0]{\begingroup\@sanitize@url \@url }%
\providecommand \@url [1]{\endgroup\@href {#1}{\urlprefix }}%
\providecommand \urlprefix  [0]{URL }%
\providecommand \Eprint [0]{\href }%
\providecommand \doibase [0]{https://doi.org/}%
\providecommand \selectlanguage [0]{\@gobble}%
\providecommand \bibinfo  [0]{\@secondoftwo}%
\providecommand \bibfield  [0]{\@secondoftwo}%
\providecommand \translation [1]{[#1]}%
\providecommand \BibitemOpen [0]{}%
\providecommand \bibitemStop [0]{}%
\providecommand \bibitemNoStop [0]{.\EOS\space}%
\providecommand \EOS [0]{\spacefactor3000\relax}%
\providecommand \BibitemShut  [1]{\csname bibitem#1\endcsname}%
\let\auto@bib@innerbib\@empty
\bibitem [{\citenamefont {Farhi}\ and\ \citenamefont
  {Jaffe}(1984)}]{Farhi-1984qu}%
  \BibitemOpen
  \bibfield  {author} {\bibinfo {author} {\bibfnamefont {E.}~\bibnamefont
  {Farhi}}\ and\ \bibinfo {author} {\bibfnamefont {R.~L.}\ \bibnamefont
  {Jaffe}},\ }\href {https://doi.org/10.1103/PhysRevD.30.2379} {\bibfield
  {journal} {\bibinfo  {journal} {Phys. Rev. D}\ }\textbf {\bibinfo {volume}
  {30}},\ \bibinfo {pages} {2379} (\bibinfo {year} {1984})}\BibitemShut
  {NoStop}%
\bibitem [{\citenamefont {Witten}(1984)}]{Witten-1984rs}%
  \BibitemOpen
  \bibfield  {author} {\bibinfo {author} {\bibfnamefont {E.}~\bibnamefont
  {Witten}},\ }\href {https://doi.org/10.1103/PhysRevD.30.272} {\bibfield
  {journal} {\bibinfo  {journal} {Phys. Rev. D}\ }\textbf {\bibinfo {volume}
  {30}},\ \bibinfo {pages} {272} (\bibinfo {year} {1984})}\BibitemShut
  {NoStop}%
\bibitem [{\citenamefont {Haensel}, \citenamefont {Zdunik},\ and\ \citenamefont
  {Schaeffer}(1986)}]{Haensel-1986qb}%
  \BibitemOpen
  \bibfield  {author} {\bibinfo {author} {\bibfnamefont {P.}~\bibnamefont
  {Haensel}}, \bibinfo {author} {\bibfnamefont {J.~L.}\ \bibnamefont
  {Zdunik}},\ and\ \bibinfo {author} {\bibfnamefont {R.}~\bibnamefont
  {Schaeffer}},\ }\href@noop {} {\bibfield  {journal} {\bibinfo  {journal}
  {Astron. Astrophys.}\ }\textbf {\bibinfo {volume} {160}},\ \bibinfo {pages}
  {121} (\bibinfo {year} {1986})}\BibitemShut {NoStop}%
\bibitem [{\citenamefont {Olinto}(1987)}]{Olinto-1986je}%
  \BibitemOpen
  \bibfield  {author} {\bibinfo {author} {\bibfnamefont {A.~V.}\ \bibnamefont
  {Olinto}},\ }\href {https://doi.org/10.1016/0370-2693(87)91144-0} {\bibfield
  {journal} {\bibinfo  {journal} {Phys. Lett. B}\ }\textbf {\bibinfo {volume}
  {192}},\ \bibinfo {pages} {71} (\bibinfo {year} {1987})}\BibitemShut
  {NoStop}%
\bibitem [{\citenamefont {Alcock}, \citenamefont {Farhi},\ and\ \citenamefont
  {Olinto}(1986)}]{Alcock-1986hz}%
  \BibitemOpen
  \bibfield  {author} {\bibinfo {author} {\bibfnamefont {C.}~\bibnamefont
  {Alcock}}, \bibinfo {author} {\bibfnamefont {E.}~\bibnamefont {Farhi}},\ and\
  \bibinfo {author} {\bibfnamefont {A.}~\bibnamefont {Olinto}},\ }\href
  {https://doi.org/10.1086/164679} {\bibfield  {journal} {\bibinfo  {journal}
  {Astrophys. J.}\ }\textbf {\bibinfo {volume} {310}},\ \bibinfo {pages} {261}
  (\bibinfo {year} {1986})}\BibitemShut {NoStop}%
\bibitem [{\citenamefont {Xu}, \citenamefont {Zhang},\ and\ \citenamefont
  {Qiao}(2001)}]{Xu-2000sr}%
  \BibitemOpen
  \bibfield  {author} {\bibinfo {author} {\bibfnamefont {R.}~\bibnamefont
  {Xu}}, \bibinfo {author} {\bibfnamefont {B.}~\bibnamefont {Zhang}},\ and\
  \bibinfo {author} {\bibfnamefont {G.}~\bibnamefont {Qiao}},\ }\href
  {https://doi.org/10.1016/S0927-6505(00)00154-7} {\bibfield  {journal}
  {\bibinfo  {journal} {Astropart. Phys.}\ }\textbf {\bibinfo {volume} {15}},\
  \bibinfo {pages} {101} (\bibinfo {year} {2001})},\ \Eprint
  {https://arxiv.org/abs/astro-ph/0006021} {arXiv:astro-ph/0006021}
  \BibitemShut {NoStop}%
\bibitem [{\citenamefont {Weber}(2005)}]{Weber-2004kj}%
  \BibitemOpen
  \bibfield  {author} {\bibinfo {author} {\bibfnamefont {F.}~\bibnamefont
  {Weber}},\ }\href {https://doi.org/10.1016/j.ppnp.2004.07.001} {\bibfield
  {journal} {\bibinfo  {journal} {Prog. Part. Nucl. Phys.}\ }\textbf {\bibinfo
  {volume} {54}},\ \bibinfo {pages} {193} (\bibinfo {year} {2005})},\ \Eprint
  {https://arxiv.org/abs/astro-ph/0407155} {arXiv:astro-ph/0407155}
  \BibitemShut {NoStop}%
\bibitem [{\citenamefont {Annala}\ \emph {et~al.}(2020)\citenamefont {Annala},
  \citenamefont {Gorda}, \citenamefont {Kurkela}, \citenamefont {N\"attil\"a},\
  and\ \citenamefont {Vuorinen}}]{Annala-2019puf}%
  \BibitemOpen
  \bibfield  {author} {\bibinfo {author} {\bibfnamefont {E.}~\bibnamefont
  {Annala}}, \bibinfo {author} {\bibfnamefont {T.}~\bibnamefont {Gorda}},
  \bibinfo {author} {\bibfnamefont {A.}~\bibnamefont {Kurkela}}, \bibinfo
  {author} {\bibfnamefont {J.}~\bibnamefont {N\"attil\"a}},\ and\ \bibinfo
  {author} {\bibfnamefont {A.}~\bibnamefont {Vuorinen}},\ }\href
  {https://doi.org/10.1038/s41567-020-0914-9} {\bibfield  {journal} {\bibinfo
  {journal} {Nature Phys.}\ }\textbf {\bibinfo {volume} {16}},\ \bibinfo
  {pages} {907} (\bibinfo {year} {2020})},\ \Eprint
  {https://arxiv.org/abs/1903.09121} {arXiv:1903.09121 [astro-ph.HE]}
  \BibitemShut {NoStop}%
\bibitem [{\citenamefont {Bombaci}\ and\ \citenamefont
  {Datta}(2000)}]{Bombaci-2000cv}%
  \BibitemOpen
  \bibfield  {author} {\bibinfo {author} {\bibfnamefont {I.}~\bibnamefont
  {Bombaci}}\ and\ \bibinfo {author} {\bibfnamefont {B.}~\bibnamefont
  {Datta}},\ }\href {https://doi.org/10.1086/312497} {\bibfield  {journal}
  {\bibinfo  {journal} {Astrophys. J. Lett.}\ }\textbf {\bibinfo {volume}
  {530}},\ \bibinfo {pages} {L69} (\bibinfo {year} {2000})},\ \Eprint
  {https://arxiv.org/abs/astro-ph/0001478} {arXiv:astro-ph/0001478}
  \BibitemShut {NoStop}%
\bibitem [{\citenamefont {Drago}\ \emph {et~al.}(2016)\citenamefont {Drago},
  \citenamefont {Lavagno}, \citenamefont {Pagliara},\ and\ \citenamefont
  {Pigato}}]{Drago-2015cea}%
  \BibitemOpen
  \bibfield  {author} {\bibinfo {author} {\bibfnamefont {A.}~\bibnamefont
  {Drago}}, \bibinfo {author} {\bibfnamefont {A.}~\bibnamefont {Lavagno}},
  \bibinfo {author} {\bibfnamefont {G.}~\bibnamefont {Pagliara}},\ and\
  \bibinfo {author} {\bibfnamefont {D.}~\bibnamefont {Pigato}},\ }\href
  {https://doi.org/10.1140/epja/i2016-16040-3} {\bibfield  {journal} {\bibinfo
  {journal} {Eur. Phys. J. A}\ }\textbf {\bibinfo {volume} {52}},\ \bibinfo
  {pages} {40} (\bibinfo {year} {2016})},\ \Eprint
  {https://arxiv.org/abs/1509.02131} {arXiv:1509.02131 [astro-ph.SR]}
  \BibitemShut {NoStop}%
\bibitem [{\citenamefont {Bhattacharyya}\ \emph {et~al.}(2017)\citenamefont
  {Bhattacharyya}, \citenamefont {Bombaci}, \citenamefont {Logoteta},\ and\
  \citenamefont {Thampan}}]{Bhattacharyya-2017mdh}%
  \BibitemOpen
  \bibfield  {author} {\bibinfo {author} {\bibfnamefont {S.}~\bibnamefont
  {Bhattacharyya}}, \bibinfo {author} {\bibfnamefont {I.}~\bibnamefont
  {Bombaci}}, \bibinfo {author} {\bibfnamefont {D.}~\bibnamefont {Logoteta}},\
  and\ \bibinfo {author} {\bibfnamefont {A.~V.}\ \bibnamefont {Thampan}},\
  }\href {https://doi.org/10.3847/1538-4357/aa8b67} {\bibfield  {journal}
  {\bibinfo  {journal} {Astrophys. J.}\ }\textbf {\bibinfo {volume} {848}},\
  \bibinfo {pages} {65} (\bibinfo {year} {2017})},\ \Eprint
  {https://arxiv.org/abs/1709.02415} {arXiv:1709.02415 [astro-ph.HE]}
  \BibitemShut {NoStop}%
\bibitem [{\citenamefont {Drago}, \citenamefont {Lavagno},\ and\ \citenamefont
  {Pagliara}(2014)}]{Drago-2013fsa}%
  \BibitemOpen
  \bibfield  {author} {\bibinfo {author} {\bibfnamefont {A.}~\bibnamefont
  {Drago}}, \bibinfo {author} {\bibfnamefont {A.}~\bibnamefont {Lavagno}},\
  and\ \bibinfo {author} {\bibfnamefont {G.}~\bibnamefont {Pagliara}},\ }\href
  {https://doi.org/10.1103/PhysRevD.89.043014} {\bibfield  {journal} {\bibinfo
  {journal} {Phys. Rev. D}\ }\textbf {\bibinfo {volume} {89}},\ \bibinfo
  {pages} {043014} (\bibinfo {year} {2014})},\ \Eprint
  {https://arxiv.org/abs/1309.7263} {arXiv:1309.7263 [nucl-th]} \BibitemShut
  {NoStop}%
\bibitem [{\citenamefont {Drago}\ and\ \citenamefont
  {Pagliara}(2016)}]{Drago-2015dea}%
  \BibitemOpen
  \bibfield  {author} {\bibinfo {author} {\bibfnamefont {A.}~\bibnamefont
  {Drago}}\ and\ \bibinfo {author} {\bibfnamefont {G.}~\bibnamefont
  {Pagliara}},\ }\href {https://doi.org/10.1140/epja/i2016-16041-2} {\bibfield
  {journal} {\bibinfo  {journal} {Eur. Phys. J. A}\ }\textbf {\bibinfo {volume}
  {52}},\ \bibinfo {pages} {41} (\bibinfo {year} {2016})},\ \Eprint
  {https://arxiv.org/abs/1509.02134} {arXiv:1509.02134 [astro-ph.SR]}
  \BibitemShut {NoStop}%
\bibitem [{\citenamefont {Bombaci}\ \emph {et~al.}(2021)\citenamefont
  {Bombaci}, \citenamefont {Drago}, \citenamefont {Logoteta}, \citenamefont
  {Pagliara},\ and\ \citenamefont {Vida\~na}}]{Bombaci-2020vgw}%
  \BibitemOpen
  \bibfield  {author} {\bibinfo {author} {\bibfnamefont {I.}~\bibnamefont
  {Bombaci}}, \bibinfo {author} {\bibfnamefont {A.}~\bibnamefont {Drago}},
  \bibinfo {author} {\bibfnamefont {D.}~\bibnamefont {Logoteta}}, \bibinfo
  {author} {\bibfnamefont {G.}~\bibnamefont {Pagliara}},\ and\ \bibinfo
  {author} {\bibfnamefont {I.}~\bibnamefont {Vida\~na}},\ }\href
  {https://doi.org/10.1103/PhysRevLett.126.162702} {\bibfield  {journal}
  {\bibinfo  {journal} {Phys. Rev. Lett.}\ }\textbf {\bibinfo {volume} {126}},\
  \bibinfo {pages} {162702} (\bibinfo {year} {2021})},\ \Eprint
  {https://arxiv.org/abs/2010.01509} {arXiv:2010.01509 [nucl-th]} \BibitemShut
  {NoStop}%
\bibitem [{\citenamefont {Drago}\ and\ \citenamefont
  {Pagliara}(2020)}]{Drago-2020gqn}%
  \BibitemOpen
  \bibfield  {author} {\bibinfo {author} {\bibfnamefont {A.}~\bibnamefont
  {Drago}}\ and\ \bibinfo {author} {\bibfnamefont {G.}~\bibnamefont
  {Pagliara}},\ }\href {https://doi.org/10.1103/PhysRevD.102.063003} {\bibfield
   {journal} {\bibinfo  {journal} {Phys. Rev. D}\ }\textbf {\bibinfo {volume}
  {102}},\ \bibinfo {pages} {063003} (\bibinfo {year} {2020})},\ \Eprint
  {https://arxiv.org/abs/2007.03436} {arXiv:2007.03436 [nucl-th]} \BibitemShut
  {NoStop}%
\bibitem [{\citenamefont {Louren\c{c}o}\ \emph {et~al.}(2021)\citenamefont
  {Louren\c{c}o}, \citenamefont {Lenzi}, \citenamefont {Dutra}, \citenamefont
  {Ferrer}, \citenamefont {de~la Incera}, \citenamefont {Paulucci},\ and\
  \citenamefont {Horvath}}]{Lourenco-2021lpn}%
  \BibitemOpen
  \bibfield  {author} {\bibinfo {author} {\bibfnamefont {O.}~\bibnamefont
  {Louren\c{c}o}}, \bibinfo {author} {\bibfnamefont {C.~H.}\ \bibnamefont
  {Lenzi}}, \bibinfo {author} {\bibfnamefont {M.}~\bibnamefont {Dutra}},
  \bibinfo {author} {\bibfnamefont {E.~J.}\ \bibnamefont {Ferrer}}, \bibinfo
  {author} {\bibfnamefont {V.}~\bibnamefont {de~la Incera}}, \bibinfo {author}
  {\bibfnamefont {L.}~\bibnamefont {Paulucci}},\ and\ \bibinfo {author}
  {\bibfnamefont {J.~E.}\ \bibnamefont {Horvath}},\ }\href
  {https://doi.org/10.1103/PhysRevD.103.103010} {\bibfield  {journal} {\bibinfo
   {journal} {Phys. Rev. D}\ }\textbf {\bibinfo {volume} {103}},\ \bibinfo
  {pages} {103010} (\bibinfo {year} {2021})},\ \Eprint
  {https://arxiv.org/abs/2104.07825} {arXiv:2104.07825 [astro-ph.HE]}
  \BibitemShut {NoStop}%
\bibitem [{\citenamefont {Haensel}, \citenamefont {Potekhin},\ and\
  \citenamefont {Yakovlev}(2007)}]{Haensel-2007yy}%
  \BibitemOpen
  \bibfield  {author} {\bibinfo {author} {\bibfnamefont {P.}~\bibnamefont
  {Haensel}}, \bibinfo {author} {\bibfnamefont {A.~Y.}\ \bibnamefont
  {Potekhin}},\ and\ \bibinfo {author} {\bibfnamefont {D.~G.}\ \bibnamefont
  {Yakovlev}},\ }\href {https://doi.org/10.1007/978-0-387-47301-7} {\emph
  {\bibinfo {title} {{Neutron stars 1: Equation of state and structure}}}},\
  Vol.\ \bibinfo {volume} {326}\ (\bibinfo  {publisher} {Springer},\ \bibinfo
  {address} {New York, USA},\ \bibinfo {year} {2007})\BibitemShut {NoStop}%
\bibitem [{\citenamefont {Duncan}\ and\ \citenamefont
  {Thompson}(1992)}]{Duncan-1992hi}%
  \BibitemOpen
  \bibfield  {author} {\bibinfo {author} {\bibfnamefont {R.~C.}\ \bibnamefont
  {Duncan}}\ and\ \bibinfo {author} {\bibfnamefont {C.}~\bibnamefont
  {Thompson}},\ }\href {https://doi.org/10.1086/186413} {\bibfield  {journal}
  {\bibinfo  {journal} {Astrophys. J.}\ }\textbf {\bibinfo {volume} {392}},\
  \bibinfo {pages} {L9} (\bibinfo {year} {1992})}\BibitemShut {NoStop}%
\bibitem [{\citenamefont {Lai}\ and\ \citenamefont
  {Shapiro}(1991)}]{91Lai-APJ}%
  \BibitemOpen
  \bibfield  {author} {\bibinfo {author} {\bibfnamefont {D.}~\bibnamefont
  {Lai}}\ and\ \bibinfo {author} {\bibfnamefont {S.~L.}\ \bibnamefont
  {Shapiro}},\ }\href {https://doi.org/10.1086/170831} {\bibfield  {journal}
  {\bibinfo  {journal} {Astrophys. J.}\ }\textbf {\bibinfo {volume} {383}},\
  \bibinfo {pages} {745} (\bibinfo {year} {1991})}\BibitemShut {NoStop}%
\bibitem [{\citenamefont {Bocquet}\ \emph {et~al.}(1995)\citenamefont
  {Bocquet}, \citenamefont {Bonazzola}, \citenamefont {Gourgoulhon},\ and\
  \citenamefont {Novak}}]{Bocquet-1995je}%
  \BibitemOpen
  \bibfield  {author} {\bibinfo {author} {\bibfnamefont {M.}~\bibnamefont
  {Bocquet}}, \bibinfo {author} {\bibfnamefont {S.}~\bibnamefont {Bonazzola}},
  \bibinfo {author} {\bibfnamefont {E.}~\bibnamefont {Gourgoulhon}},\ and\
  \bibinfo {author} {\bibfnamefont {J.}~\bibnamefont {Novak}},\ }\href@noop {}
  {\bibfield  {journal} {\bibinfo  {journal} {Astron. Astrophys.}\ }\textbf
  {\bibinfo {volume} {301}},\ \bibinfo {pages} {757} (\bibinfo {year}
  {1995})},\ \Eprint {https://arxiv.org/abs/gr-qc/9503044}
  {arXiv:gr-qc/9503044} \BibitemShut {NoStop}%
\bibitem [{\citenamefont {Broderick}, \citenamefont {Prakash},\ and\
  \citenamefont {Lattimer}(2002)}]{Broderick-2001qw}%
  \BibitemOpen
  \bibfield  {author} {\bibinfo {author} {\bibfnamefont {A.~E.}\ \bibnamefont
  {Broderick}}, \bibinfo {author} {\bibfnamefont {M.}~\bibnamefont {Prakash}},\
  and\ \bibinfo {author} {\bibfnamefont {J.~M.}\ \bibnamefont {Lattimer}},\
  }\href {https://doi.org/10.1016/S0370-2693(01)01514-3} {\bibfield  {journal}
  {\bibinfo  {journal} {Phys. Lett. B}\ }\textbf {\bibinfo {volume} {531}},\
  \bibinfo {pages} {167} (\bibinfo {year} {2002})},\ \Eprint
  {https://arxiv.org/abs/astro-ph/0111516} {arXiv:astro-ph/0111516}
  \BibitemShut {NoStop}%
\bibitem [{\citenamefont {Ferrer}\ \emph {et~al.}(2010)\citenamefont {Ferrer},
  \citenamefont {de~la Incera}, \citenamefont {Keith}, \citenamefont
  {Portillo},\ and\ \citenamefont {Springsteen}}]{Ferrer-2010wz}%
  \BibitemOpen
  \bibfield  {author} {\bibinfo {author} {\bibfnamefont {E.~J.}\ \bibnamefont
  {Ferrer}}, \bibinfo {author} {\bibfnamefont {V.}~\bibnamefont {de~la
  Incera}}, \bibinfo {author} {\bibfnamefont {J.~P.}\ \bibnamefont {Keith}},
  \bibinfo {author} {\bibfnamefont {I.}~\bibnamefont {Portillo}},\ and\
  \bibinfo {author} {\bibfnamefont {P.~L.}\ \bibnamefont {Springsteen}},\
  }\href {https://doi.org/10.1103/PhysRevC.82.065802} {\bibfield  {journal}
  {\bibinfo  {journal} {Phys. Rev. C}\ }\textbf {\bibinfo {volume} {82}},\
  \bibinfo {pages} {065802} (\bibinfo {year} {2010})},\ \Eprint
  {https://arxiv.org/abs/1009.3521} {arXiv:1009.3521 [hep-ph]} \BibitemShut
  {NoStop}%
\bibitem [{\citenamefont {Ferrer}\ and\ \citenamefont {de~la
  Incera}(2013)}]{Ferrer-2012wa}%
  \BibitemOpen
  \bibfield  {author} {\bibinfo {author} {\bibfnamefont {E.~J.}\ \bibnamefont
  {Ferrer}}\ and\ \bibinfo {author} {\bibfnamefont {V.}~\bibnamefont {de~la
  Incera}},\ }\href {https://doi.org/10.1007/978-3-642-37305-3_16} {\bibfield
  {journal} {\bibinfo  {journal} {Lect. Notes Phys.}\ }\textbf {\bibinfo
  {volume} {871}},\ \bibinfo {pages} {399} (\bibinfo {year} {2013})},\ \Eprint
  {https://arxiv.org/abs/1208.5179} {arXiv:1208.5179 [nucl-th]} \BibitemShut
  {NoStop}%
\bibitem [{\citenamefont {Kharzeev}, \citenamefont {McLerran},\ and\
  \citenamefont {Warringa}(2008)}]{08Kharzeev.McLerran.ea227-253NPA}%
  \BibitemOpen
  \bibfield  {author} {\bibinfo {author} {\bibfnamefont {D.~E.}\ \bibnamefont
  {Kharzeev}}, \bibinfo {author} {\bibfnamefont {L.~D.}\ \bibnamefont
  {McLerran}},\ and\ \bibinfo {author} {\bibfnamefont {H.~J.}\ \bibnamefont
  {Warringa}},\ }\href {https://doi.org/10.1016/j.nuclphysa.2008.02.298}
  {\bibfield  {journal} {\bibinfo  {journal} {Nucl. Phys. A}\ }\textbf
  {\bibinfo {volume} {803}},\ \bibinfo {pages} {227} (\bibinfo {year}
  {2008})},\ \Eprint {https://arxiv.org/abs/0711.0950} {arXiv:0711.0950
  [hep-ph]} \BibitemShut {NoStop}%
\bibitem [{\citenamefont {Fukushima}, \citenamefont {Kharzeev},\ and\
  \citenamefont {Warringa}(2008)}]{Fukushima-2008xe}%
  \BibitemOpen
  \bibfield  {author} {\bibinfo {author} {\bibfnamefont {K.}~\bibnamefont
  {Fukushima}}, \bibinfo {author} {\bibfnamefont {D.~E.}\ \bibnamefont
  {Kharzeev}},\ and\ \bibinfo {author} {\bibfnamefont {H.~J.}\ \bibnamefont
  {Warringa}},\ }\href {https://doi.org/10.1103/PhysRevD.78.074033} {\bibfield
  {journal} {\bibinfo  {journal} {Phys. Rev. D}\ }\textbf {\bibinfo {volume}
  {78}},\ \bibinfo {pages} {074033} (\bibinfo {year} {2008})},\ \Eprint
  {https://arxiv.org/abs/0808.3382} {arXiv:0808.3382 [hep-ph]} \BibitemShut
  {NoStop}%
\bibitem [{\citenamefont {Skokov}, \citenamefont {Illarionov},\ and\
  \citenamefont {Toneev}(2009)}]{Skokov-2009qp}%
  \BibitemOpen
  \bibfield  {author} {\bibinfo {author} {\bibfnamefont {V.}~\bibnamefont
  {Skokov}}, \bibinfo {author} {\bibfnamefont {A.}~\bibnamefont {Illarionov}},\
  and\ \bibinfo {author} {\bibfnamefont {V.}~\bibnamefont {Toneev}},\ }\href
  {https://doi.org/10.1142/S0217751X09047570} {\bibfield  {journal} {\bibinfo
  {journal} {Int. J. Mod. Phys. A}\ }\textbf {\bibinfo {volume} {24}},\
  \bibinfo {pages} {5925} (\bibinfo {year} {2009})},\ \Eprint
  {https://arxiv.org/abs/0907.1396} {arXiv:0907.1396 [nucl-th]} \BibitemShut
  {NoStop}%
\bibitem [{\citenamefont {Voronyuk}\ \emph {et~al.}(2011)\citenamefont
  {Voronyuk}, \citenamefont {Toneev}, \citenamefont {Cassing}, \citenamefont
  {Bratkovskaya}, \citenamefont {Konchakovski},\ and\ \citenamefont
  {Voloshin}}]{Voronyuk-2011jd}%
  \BibitemOpen
  \bibfield  {author} {\bibinfo {author} {\bibfnamefont {V.}~\bibnamefont
  {Voronyuk}}, \bibinfo {author} {\bibfnamefont {V.}~\bibnamefont {Toneev}},
  \bibinfo {author} {\bibfnamefont {W.}~\bibnamefont {Cassing}}, \bibinfo
  {author} {\bibfnamefont {E.}~\bibnamefont {Bratkovskaya}}, \bibinfo {author}
  {\bibfnamefont {V.}~\bibnamefont {Konchakovski}},\ and\ \bibinfo {author}
  {\bibfnamefont {S.}~\bibnamefont {Voloshin}},\ }\href
  {https://doi.org/10.1103/PhysRevC.83.054911} {\bibfield  {journal} {\bibinfo
  {journal} {Phys. Rev. C}\ }\textbf {\bibinfo {volume} {83}},\ \bibinfo
  {pages} {054911} (\bibinfo {year} {2011})},\ \Eprint
  {https://arxiv.org/abs/1103.4239} {arXiv:1103.4239 [nucl-th]} \BibitemShut
  {NoStop}%
\bibitem [{\citenamefont {Deng}\ and\ \citenamefont
  {Huang}(2012)}]{Deng-2012pc}%
  \BibitemOpen
  \bibfield  {author} {\bibinfo {author} {\bibfnamefont {W.-T.}\ \bibnamefont
  {Deng}}\ and\ \bibinfo {author} {\bibfnamefont {X.-G.}\ \bibnamefont
  {Huang}},\ }\href {https://doi.org/10.1103/PhysRevC.85.044907} {\bibfield
  {journal} {\bibinfo  {journal} {Phys. Rev. C}\ }\textbf {\bibinfo {volume}
  {85}},\ \bibinfo {pages} {044907} (\bibinfo {year} {2012})},\ \Eprint
  {https://arxiv.org/abs/1201.5108} {arXiv:1201.5108 [nucl-th]} \BibitemShut
  {NoStop}%
\bibitem [{\citenamefont {Chatterjee}\ \emph {et~al.}(2015)\citenamefont
  {Chatterjee}, \citenamefont {Elghozi}, \citenamefont {Novak},\ and\
  \citenamefont {Oertel}}]{Chatterjee-2014qsa}%
  \BibitemOpen
  \bibfield  {author} {\bibinfo {author} {\bibfnamefont {D.}~\bibnamefont
  {Chatterjee}}, \bibinfo {author} {\bibfnamefont {T.}~\bibnamefont {Elghozi}},
  \bibinfo {author} {\bibfnamefont {J.}~\bibnamefont {Novak}},\ and\ \bibinfo
  {author} {\bibfnamefont {M.}~\bibnamefont {Oertel}},\ }\href
  {https://doi.org/10.1093/mnras/stu2706} {\bibfield  {journal} {\bibinfo
  {journal} {Mon. Not. Roy. Astron. Soc.}\ }\textbf {\bibinfo {volume} {447}},\
  \bibinfo {pages} {3785} (\bibinfo {year} {2015})},\ \Eprint
  {https://arxiv.org/abs/1410.6332} {arXiv:1410.6332 [astro-ph.HE]}
  \BibitemShut {NoStop}%
\bibitem [{\citenamefont {Chakrabarty}(1996)}]{Chakrabarty-1996te}%
  \BibitemOpen
  \bibfield  {author} {\bibinfo {author} {\bibfnamefont {S.}~\bibnamefont
  {Chakrabarty}},\ }\href {https://doi.org/10.1103/PhysRevD.54.1306} {\bibfield
   {journal} {\bibinfo  {journal} {Phys. Rev. D}\ }\textbf {\bibinfo {volume}
  {54}},\ \bibinfo {pages} {1306} (\bibinfo {year} {1996})},\ \Eprint
  {https://arxiv.org/abs/hep-ph/9603406} {arXiv:hep-ph/9603406} \BibitemShut
  {NoStop}%
\bibitem [{\citenamefont {Isayev}(2015)}]{Isayev-2015jqa}%
  \BibitemOpen
  \bibfield  {author} {\bibinfo {author} {\bibfnamefont {A.~A.}\ \bibnamefont
  {Isayev}},\ }\href {https://doi.org/10.1103/PhysRevC.91.015208} {\bibfield
  {journal} {\bibinfo  {journal} {Phys. Rev. C}\ }\textbf {\bibinfo {volume}
  {91}},\ \bibinfo {pages} {015208} (\bibinfo {year} {2015})},\ \Eprint
  {https://arxiv.org/abs/1501.07772} {arXiv:1501.07772 [hep-ph]} \BibitemShut
  {NoStop}%
\bibitem [{\citenamefont {Menezes}\ \emph
  {et~al.}(2009{\natexlab{a}})\citenamefont {Menezes}, \citenamefont
  {Benghi~Pinto}, \citenamefont {Avancini},\ and\ \citenamefont
  {Providencia}}]{Menezes-2009uc}%
  \BibitemOpen
  \bibfield  {author} {\bibinfo {author} {\bibfnamefont {D.~P.}\ \bibnamefont
  {Menezes}}, \bibinfo {author} {\bibfnamefont {M.}~\bibnamefont
  {Benghi~Pinto}}, \bibinfo {author} {\bibfnamefont {S.~S.}\ \bibnamefont
  {Avancini}},\ and\ \bibinfo {author} {\bibfnamefont {C.}~\bibnamefont
  {Providencia}},\ }\href {https://doi.org/10.1103/PhysRevC.80.065805}
  {\bibfield  {journal} {\bibinfo  {journal} {Phys. Rev. C}\ }\textbf {\bibinfo
  {volume} {80}},\ \bibinfo {pages} {065805} (\bibinfo {year}
  {2009}{\natexlab{a}})},\ \Eprint {https://arxiv.org/abs/0907.2607}
  {arXiv:0907.2607 [nucl-th]} \BibitemShut {NoStop}%
\bibitem [{\citenamefont {Chu}\ \emph {et~al.}(2018)\citenamefont {Chu},
  \citenamefont {Li}, \citenamefont {Ma}, \citenamefont {Wang}, \citenamefont
  {Dong},\ and\ \citenamefont {Zhang}}]{Chu-2018dch}%
  \BibitemOpen
  \bibfield  {author} {\bibinfo {author} {\bibfnamefont {P.-C.}\ \bibnamefont
  {Chu}}, \bibinfo {author} {\bibfnamefont {X.-H.}\ \bibnamefont {Li}},
  \bibinfo {author} {\bibfnamefont {H.-Y.}\ \bibnamefont {Ma}}, \bibinfo
  {author} {\bibfnamefont {B.}~\bibnamefont {Wang}}, \bibinfo {author}
  {\bibfnamefont {Y.-M.}\ \bibnamefont {Dong}},\ and\ \bibinfo {author}
  {\bibfnamefont {X.-M.}\ \bibnamefont {Zhang}},\ }\href
  {https://doi.org/10.1016/j.physletb.2018.01.064} {\bibfield  {journal}
  {\bibinfo  {journal} {Phys. Lett. B}\ }\textbf {\bibinfo {volume} {778}},\
  \bibinfo {pages} {447} (\bibinfo {year} {2018})}\BibitemShut {NoStop}%
\bibitem [{\citenamefont {Chu}\ \emph {et~al.}(2020)\citenamefont {Chu},
  \citenamefont {Zhou}, \citenamefont {Chen}, \citenamefont {Li},\ and\
  \citenamefont {Ma}}]{Chu-2020afg}%
  \BibitemOpen
  \bibfield  {author} {\bibinfo {author} {\bibfnamefont {P.-C.}\ \bibnamefont
  {Chu}}, \bibinfo {author} {\bibfnamefont {Y.}~\bibnamefont {Zhou}}, \bibinfo
  {author} {\bibfnamefont {C.}~\bibnamefont {Chen}}, \bibinfo {author}
  {\bibfnamefont {X.-H.}\ \bibnamefont {Li}},\ and\ \bibinfo {author}
  {\bibfnamefont {H.-Y.}\ \bibnamefont {Ma}},\ }\href
  {https://doi.org/10.1088/1361-6471/ab9b06} {\bibfield  {journal} {\bibinfo
  {journal} {J. Phys. G}\ }\textbf {\bibinfo {volume} {47}},\ \bibinfo {pages}
  {085201} (\bibinfo {year} {2020})}\BibitemShut {NoStop}%
\bibitem [{\citenamefont {Isayev}(2018)}]{Isayev-2018hzq}%
  \BibitemOpen
  \bibfield  {author} {\bibinfo {author} {\bibfnamefont {A.~A.}\ \bibnamefont
  {Isayev}},\ }\href {https://doi.org/10.1103/PhysRevD.98.043022} {\bibfield
  {journal} {\bibinfo  {journal} {Phys. Rev. D}\ }\textbf {\bibinfo {volume}
  {98}},\ \bibinfo {pages} {043022} (\bibinfo {year} {2018})},\ \Eprint
  {https://arxiv.org/abs/1809.00713} {arXiv:1809.00713 [hep-ph]} \BibitemShut
  {NoStop}%
\bibitem [{\citenamefont {Lugones}\ and\ \citenamefont
  {Grunfeld}(2019)}]{Lugones-2018qgu}%
  \BibitemOpen
  \bibfield  {author} {\bibinfo {author} {\bibfnamefont {G.}~\bibnamefont
  {Lugones}}\ and\ \bibinfo {author} {\bibfnamefont {A.~G.}\ \bibnamefont
  {Grunfeld}},\ }\href {https://doi.org/10.1103/PhysRevC.99.035804} {\bibfield
  {journal} {\bibinfo  {journal} {Phys. Rev. C}\ }\textbf {\bibinfo {volume}
  {99}},\ \bibinfo {pages} {035804} (\bibinfo {year} {2019})},\ \Eprint
  {https://arxiv.org/abs/1811.09954} {arXiv:1811.09954 [astro-ph.HE]}
  \BibitemShut {NoStop}%
\bibitem [{\citenamefont {Huang}\ \emph {et~al.}(2010)\citenamefont {Huang},
  \citenamefont {Huang}, \citenamefont {Rischke},\ and\ \citenamefont
  {Sedrakian}}]{Huang-2009ue}%
  \BibitemOpen
  \bibfield  {author} {\bibinfo {author} {\bibfnamefont {X.-G.}\ \bibnamefont
  {Huang}}, \bibinfo {author} {\bibfnamefont {M.}~\bibnamefont {Huang}},
  \bibinfo {author} {\bibfnamefont {D.~H.}\ \bibnamefont {Rischke}},\ and\
  \bibinfo {author} {\bibfnamefont {A.}~\bibnamefont {Sedrakian}},\ }\href
  {https://doi.org/10.1103/PhysRevD.81.045015} {\bibfield  {journal} {\bibinfo
  {journal} {Phys. Rev. D}\ }\textbf {\bibinfo {volume} {81}},\ \bibinfo
  {pages} {045015} (\bibinfo {year} {2010})},\ \Eprint
  {https://arxiv.org/abs/0910.3633} {arXiv:0910.3633 [astro-ph.HE]}
  \BibitemShut {NoStop}%
\bibitem [{\citenamefont {Xu}\ \emph {et~al.}(2021)\citenamefont {Xu},
  \citenamefont {Kang}, \citenamefont {Peng},\ and\ \citenamefont
  {Xia}}]{Xu-2021eyi}%
  \BibitemOpen
  \bibfield  {author} {\bibinfo {author} {\bibfnamefont {J.-F.}\ \bibnamefont
  {Xu}}, \bibinfo {author} {\bibfnamefont {D.-B.}\ \bibnamefont {Kang}},
  \bibinfo {author} {\bibfnamefont {G.-X.}\ \bibnamefont {Peng}},\ and\
  \bibinfo {author} {\bibfnamefont {C.-J.}\ \bibnamefont {Xia}},\ }\href
  {https://doi.org/10.1088/1674-1137/abc0cd} {\bibfield  {journal} {\bibinfo
  {journal} {Chin. Phys. C}\ }\textbf {\bibinfo {volume} {45}},\ \bibinfo
  {pages} {015103} (\bibinfo {year} {2021})}\BibitemShut {NoStop}%
\bibitem [{\citenamefont {Hou}\ \emph {et~al.}(2015)\citenamefont {Hou},
  \citenamefont {Peng}, \citenamefont {Xia},\ and\ \citenamefont
  {Xu}}]{Hou-2014pba}%
  \BibitemOpen
  \bibfield  {author} {\bibinfo {author} {\bibfnamefont {J.-X.}\ \bibnamefont
  {Hou}}, \bibinfo {author} {\bibfnamefont {G.-X.}\ \bibnamefont {Peng}},
  \bibinfo {author} {\bibfnamefont {C.-J.}\ \bibnamefont {Xia}},\ and\ \bibinfo
  {author} {\bibfnamefont {J.-F.}\ \bibnamefont {Xu}},\ }\href
  {https://doi.org/10.1088/1674-1137/39/1/015101} {\bibfield  {journal}
  {\bibinfo  {journal} {Chin. Phys. C}\ }\textbf {\bibinfo {volume} {39}},\
  \bibinfo {pages} {015101} (\bibinfo {year} {2015})},\ \Eprint
  {https://arxiv.org/abs/1403.1143} {arXiv:1403.1143 [nucl-th]} \BibitemShut
  {NoStop}%
\bibitem [{\citenamefont {Peres~Menezes}\ and\ \citenamefont
  {La\'ercio~Lopes}(2016)}]{PeresMenezes-2015ukv}%
  \BibitemOpen
  \bibfield  {author} {\bibinfo {author} {\bibfnamefont {D.}~\bibnamefont
  {Peres~Menezes}}\ and\ \bibinfo {author} {\bibfnamefont {L.}~\bibnamefont
  {La\'ercio~Lopes}},\ }\href {https://doi.org/10.1140/epja/i2016-16017-2}
  {\bibfield  {journal} {\bibinfo  {journal} {Eur. Phys. J. A}\ }\textbf
  {\bibinfo {volume} {52}},\ \bibinfo {pages} {17} (\bibinfo {year} {2016})},\
  \Eprint {https://arxiv.org/abs/1505.06714} {arXiv:1505.06714 [nucl-th]}
  \BibitemShut {NoStop}%
\bibitem [{\citenamefont {Chu}, \citenamefont {Chen},\ and\ \citenamefont
  {Wang}(2014)}]{Chu-2014foa}%
  \BibitemOpen
  \bibfield  {author} {\bibinfo {author} {\bibfnamefont {P.-C.}\ \bibnamefont
  {Chu}}, \bibinfo {author} {\bibfnamefont {L.-W.}\ \bibnamefont {Chen}},\ and\
  \bibinfo {author} {\bibfnamefont {X.}~\bibnamefont {Wang}},\ }\href
  {https://doi.org/10.1103/PhysRevD.90.063013} {\bibfield  {journal} {\bibinfo
  {journal} {Phys. Rev. D}\ }\textbf {\bibinfo {volume} {90}},\ \bibinfo
  {pages} {063013} (\bibinfo {year} {2014})},\ \Eprint
  {https://arxiv.org/abs/1406.5610} {arXiv:1406.5610 [nucl-th]} \BibitemShut
  {NoStop}%
\bibitem [{\citenamefont {Kayanikhoo}, \citenamefont {Naficy},\ and\
  \citenamefont {Bordbar}(2020)}]{Kayanikhoo-2019ugo}%
  \BibitemOpen
  \bibfield  {author} {\bibinfo {author} {\bibfnamefont {F.}~\bibnamefont
  {Kayanikhoo}}, \bibinfo {author} {\bibfnamefont {K.}~\bibnamefont {Naficy}},\
  and\ \bibinfo {author} {\bibfnamefont {G.~H.}\ \bibnamefont {Bordbar}},\
  }\href {https://doi.org/10.1140/epja/s10050-019-00004-y} {\bibfield
  {journal} {\bibinfo  {journal} {Eur. Phys. J. A}\ }\textbf {\bibinfo {volume}
  {56}},\ \bibinfo {pages} {2} (\bibinfo {year} {2020})},\ \Eprint
  {https://arxiv.org/abs/1911.10512} {arXiv:1911.10512 [nucl-th]} \BibitemShut
  {NoStop}%
\bibitem [{\citenamefont {Chu}\ \emph {et~al.}(2021)\citenamefont {Chu},
  \citenamefont {Li}, \citenamefont {Liu},\ and\ \citenamefont
  {Zhang}}]{Chu-2021oey}%
  \BibitemOpen
  \bibfield  {author} {\bibinfo {author} {\bibfnamefont {P.-C.}\ \bibnamefont
  {Chu}}, \bibinfo {author} {\bibfnamefont {X.-H.}\ \bibnamefont {Li}},
  \bibinfo {author} {\bibfnamefont {H.}~\bibnamefont {Liu}},\ and\ \bibinfo
  {author} {\bibfnamefont {J.-W.}\ \bibnamefont {Zhang}},\ }\href
  {https://doi.org/10.1103/PhysRevC.104.045805} {\bibfield  {journal} {\bibinfo
   {journal} {Phys. Rev. C}\ }\textbf {\bibinfo {volume} {104}},\ \bibinfo
  {pages} {045805} (\bibinfo {year} {2021})}\BibitemShut {NoStop}%
\bibitem [{\citenamefont {Menezes}\ \emph
  {et~al.}(2009{\natexlab{b}})\citenamefont {Menezes}, \citenamefont
  {Benghi~Pinto}, \citenamefont {Avancini}, \citenamefont {Perez~Martinez},\
  and\ \citenamefont {Providencia}}]{Menezes-2008qt}%
  \BibitemOpen
  \bibfield  {author} {\bibinfo {author} {\bibfnamefont {D.~P.}\ \bibnamefont
  {Menezes}}, \bibinfo {author} {\bibfnamefont {M.}~\bibnamefont
  {Benghi~Pinto}}, \bibinfo {author} {\bibfnamefont {S.~S.}\ \bibnamefont
  {Avancini}}, \bibinfo {author} {\bibfnamefont {A.}~\bibnamefont
  {Perez~Martinez}},\ and\ \bibinfo {author} {\bibfnamefont {C.}~\bibnamefont
  {Providencia}},\ }\href {https://doi.org/10.1103/PhysRevC.79.035807}
  {\bibfield  {journal} {\bibinfo  {journal} {Phys. Rev. C}\ }\textbf {\bibinfo
  {volume} {79}},\ \bibinfo {pages} {035807} (\bibinfo {year}
  {2009}{\natexlab{b}})},\ \Eprint {https://arxiv.org/abs/0811.3361}
  {arXiv:0811.3361 [nucl-th]} \BibitemShut {NoStop}%
\bibitem [{\citenamefont {Ruggieri}, \citenamefont {Lu},\ and\ \citenamefont
  {Peng}(2016)}]{Ruggieri-2016xww}%
  \BibitemOpen
  \bibfield  {author} {\bibinfo {author} {\bibfnamefont {M.}~\bibnamefont
  {Ruggieri}}, \bibinfo {author} {\bibfnamefont {Z.-Y.}\ \bibnamefont {Lu}},\
  and\ \bibinfo {author} {\bibfnamefont {G.-X.}\ \bibnamefont {Peng}},\ }\href
  {https://doi.org/10.1103/PhysRevD.94.116003} {\bibfield  {journal} {\bibinfo
  {journal} {Phys. Rev. D}\ }\textbf {\bibinfo {volume} {94}},\ \bibinfo
  {pages} {116003} (\bibinfo {year} {2016})},\ \Eprint
  {https://arxiv.org/abs/1608.08310} {arXiv:1608.08310 [hep-ph]} \BibitemShut
  {NoStop}%
\bibitem [{\citenamefont {Wen}, \citenamefont {He},\ and\ \citenamefont
  {Liu}(2021)}]{Wen-2021mgm}%
  \BibitemOpen
  \bibfield  {author} {\bibinfo {author} {\bibfnamefont {X.-J.}\ \bibnamefont
  {Wen}}, \bibinfo {author} {\bibfnamefont {R.}~\bibnamefont {He}},\ and\
  \bibinfo {author} {\bibfnamefont {J.-B.}\ \bibnamefont {Liu}},\ }\href
  {https://doi.org/10.1103/PhysRevD.103.094020} {\bibfield  {journal} {\bibinfo
   {journal} {Phys. Rev. D}\ }\textbf {\bibinfo {volume} {103}},\ \bibinfo
  {pages} {094020} (\bibinfo {year} {2021})}\BibitemShut {NoStop}%
\bibitem [{\citenamefont {Ferrer}\ and\ \citenamefont
  {Hackebill}(2020)}]{Ferrer-2020tlz}%
  \BibitemOpen
  \bibfield  {author} {\bibinfo {author} {\bibfnamefont {E.~J.}\ \bibnamefont
  {Ferrer}}\ and\ \bibinfo {author} {\bibfnamefont {A.}~\bibnamefont
  {Hackebill}},\ }\href@noop {} {\  (\bibinfo {year} {2020})},\ \Eprint
  {https://arxiv.org/abs/2010.10574} {arXiv:2010.10574 [nucl-th]} \BibitemShut
  {NoStop}%
\bibitem [{\citenamefont {Xu}, \citenamefont {Chao},\ and\ \citenamefont
  {Huang}(2021)}]{Xu-2020yag}%
  \BibitemOpen
  \bibfield  {author} {\bibinfo {author} {\bibfnamefont {K.}~\bibnamefont
  {Xu}}, \bibinfo {author} {\bibfnamefont {J.}~\bibnamefont {Chao}},\ and\
  \bibinfo {author} {\bibfnamefont {M.}~\bibnamefont {Huang}},\ }\href
  {https://doi.org/10.1103/PhysRevD.103.076015} {\bibfield  {journal} {\bibinfo
   {journal} {Phys. Rev. D}\ }\textbf {\bibinfo {volume} {103}},\ \bibinfo
  {pages} {076015} (\bibinfo {year} {2021})},\ \Eprint
  {https://arxiv.org/abs/2007.13122} {arXiv:2007.13122 [hep-ph]} \BibitemShut
  {NoStop}%
\bibitem [{\citenamefont {Chaudhuri}\ \emph {et~al.}(2019)\citenamefont
  {Chaudhuri}, \citenamefont {Ghosh}, \citenamefont {Sarkar},\ and\
  \citenamefont {Roy}}]{Chaudhuri-2019lbw}%
  \BibitemOpen
  \bibfield  {author} {\bibinfo {author} {\bibfnamefont {N.}~\bibnamefont
  {Chaudhuri}}, \bibinfo {author} {\bibfnamefont {S.}~\bibnamefont {Ghosh}},
  \bibinfo {author} {\bibfnamefont {S.}~\bibnamefont {Sarkar}},\ and\ \bibinfo
  {author} {\bibfnamefont {P.}~\bibnamefont {Roy}},\ }\href
  {https://doi.org/10.1103/PhysRevD.99.116025} {\bibfield  {journal} {\bibinfo
  {journal} {Phys. Rev. D}\ }\textbf {\bibinfo {volume} {99}},\ \bibinfo
  {pages} {116025} (\bibinfo {year} {2019})},\ \Eprint
  {https://arxiv.org/abs/1907.03990} {arXiv:1907.03990 [nucl-th]} \BibitemShut
  {NoStop}%
\bibitem [{\citenamefont {Chaudhuri}\ \emph {et~al.}(2020)\citenamefont
  {Chaudhuri}, \citenamefont {Ghosh}, \citenamefont {Sarkar},\ and\
  \citenamefont {Roy}}]{Chaudhuri-2020lga}%
  \BibitemOpen
  \bibfield  {author} {\bibinfo {author} {\bibfnamefont {N.}~\bibnamefont
  {Chaudhuri}}, \bibinfo {author} {\bibfnamefont {S.}~\bibnamefont {Ghosh}},
  \bibinfo {author} {\bibfnamefont {S.}~\bibnamefont {Sarkar}},\ and\ \bibinfo
  {author} {\bibfnamefont {P.}~\bibnamefont {Roy}},\ }\href
  {https://doi.org/10.1140/epja/s10050-020-00222-9} {\bibfield  {journal}
  {\bibinfo  {journal} {Eur. Phys. J. A}\ }\textbf {\bibinfo {volume} {56}},\
  \bibinfo {pages} {213} (\bibinfo {year} {2020})},\ \Eprint
  {https://arxiv.org/abs/2003.05692} {arXiv:2003.05692 [nucl-th]} \BibitemShut
  {NoStop}%
\bibitem [{\citenamefont {Grunfeld}\ \emph {et~al.}(2014)\citenamefont
  {Grunfeld}, \citenamefont {Menezes}, \citenamefont {Pinto},\ and\
  \citenamefont {Scoccola}}]{Grunfeld-2014qfa}%
  \BibitemOpen
  \bibfield  {author} {\bibinfo {author} {\bibfnamefont {A.}~\bibnamefont
  {Grunfeld}}, \bibinfo {author} {\bibfnamefont {D.}~\bibnamefont {Menezes}},
  \bibinfo {author} {\bibfnamefont {M.}~\bibnamefont {Pinto}},\ and\ \bibinfo
  {author} {\bibfnamefont {N.}~\bibnamefont {Scoccola}},\ }\href
  {https://doi.org/10.1103/PhysRevD.90.044024} {\bibfield  {journal} {\bibinfo
  {journal} {Phys. Rev. D}\ }\textbf {\bibinfo {volume} {90}},\ \bibinfo
  {pages} {044024} (\bibinfo {year} {2014})},\ \Eprint
  {https://arxiv.org/abs/1402.4731} {arXiv:1402.4731 [hep-ph]} \BibitemShut
  {NoStop}%
\bibitem [{\citenamefont {Farias}\ \emph {et~al.}(2021)\citenamefont {Farias},
  \citenamefont {Tavares}, \citenamefont {Nunes},\ and\ \citenamefont
  {Avancini}}]{Farias-2021fci}%
  \BibitemOpen
  \bibfield  {author} {\bibinfo {author} {\bibfnamefont {R.~L.~S.}\
  \bibnamefont {Farias}}, \bibinfo {author} {\bibfnamefont {W.~R.}\
  \bibnamefont {Tavares}}, \bibinfo {author} {\bibfnamefont {R.~M.}\
  \bibnamefont {Nunes}},\ and\ \bibinfo {author} {\bibfnamefont {S.~S.}\
  \bibnamefont {Avancini}},\ }\href@noop {} {\  (\bibinfo {year} {2021})},\
  \Eprint {https://arxiv.org/abs/2109.11112} {arXiv:2109.11112 [hep-ph]}
  \BibitemShut {NoStop}%
\bibitem [{\citenamefont {Fraga}\ and\ \citenamefont
  {Palhares}(2012)}]{Fraga-2012fs}%
  \BibitemOpen
  \bibfield  {author} {\bibinfo {author} {\bibfnamefont {E.~S.}\ \bibnamefont
  {Fraga}}\ and\ \bibinfo {author} {\bibfnamefont {L.~F.}\ \bibnamefont
  {Palhares}},\ }\href {https://doi.org/10.1103/PhysRevD.86.016008} {\bibfield
  {journal} {\bibinfo  {journal} {Phys. Rev. D}\ }\textbf {\bibinfo {volume}
  {86}},\ \bibinfo {pages} {016008} (\bibinfo {year} {2012})},\ \Eprint
  {https://arxiv.org/abs/1201.5881} {arXiv:1201.5881 [hep-ph]} \BibitemShut
  {NoStop}%
\bibitem [{\citenamefont {Mizher}, \citenamefont {Chernodub},\ and\
  \citenamefont {Fraga}(2010)}]{Mizher-2010zb}%
  \BibitemOpen
  \bibfield  {author} {\bibinfo {author} {\bibfnamefont {A.~J.}\ \bibnamefont
  {Mizher}}, \bibinfo {author} {\bibfnamefont {M.~N.}\ \bibnamefont
  {Chernodub}},\ and\ \bibinfo {author} {\bibfnamefont {E.~S.}\ \bibnamefont
  {Fraga}},\ }\href {https://doi.org/10.1103/PhysRevD.82.105016} {\bibfield
  {journal} {\bibinfo  {journal} {Phys. Rev. D}\ }\textbf {\bibinfo {volume}
  {82}},\ \bibinfo {pages} {105016} (\bibinfo {year} {2010})},\ \Eprint
  {https://arxiv.org/abs/1004.2712} {arXiv:1004.2712 [hep-ph]} \BibitemShut
  {NoStop}%
\bibitem [{\citenamefont {Ferreira}\ \emph {et~al.}(2014)\citenamefont
  {Ferreira}, \citenamefont {Costa}, \citenamefont {Menezes}, \citenamefont
  {Provid\^encia},\ and\ \citenamefont {Scoccola}}]{Ferreira-2013tba}%
  \BibitemOpen
  \bibfield  {author} {\bibinfo {author} {\bibfnamefont {M.}~\bibnamefont
  {Ferreira}}, \bibinfo {author} {\bibfnamefont {P.}~\bibnamefont {Costa}},
  \bibinfo {author} {\bibfnamefont {D.~P.}\ \bibnamefont {Menezes}}, \bibinfo
  {author} {\bibfnamefont {C.}~\bibnamefont {Provid\^encia}},\ and\ \bibinfo
  {author} {\bibfnamefont {N.}~\bibnamefont {Scoccola}},\ }\href
  {https://doi.org/10.1103/PhysRevD.89.016002} {\bibfield  {journal} {\bibinfo
  {journal} {Phys. Rev. D}\ }\textbf {\bibinfo {volume} {89}},\ \bibinfo
  {pages} {016002} (\bibinfo {year} {2014})},\ \bibinfo {note} {[Addendum:
  Phys.Rev.D 89, 019902 (2014)]},\ \Eprint {https://arxiv.org/abs/1305.4751}
  {arXiv:1305.4751 [hep-ph]} \BibitemShut {NoStop}%
\bibitem [{\citenamefont {Gatto}\ and\ \citenamefont
  {Ruggieri}(2011)}]{11Gatto.Ruggieri34016-34016PRD}%
  \BibitemOpen
  \bibfield  {author} {\bibinfo {author} {\bibfnamefont {R.}~\bibnamefont
  {Gatto}}\ and\ \bibinfo {author} {\bibfnamefont {M.}~\bibnamefont
  {Ruggieri}},\ }\href {https://doi.org/10.1103/PhysRevD.83.034016} {\bibfield
  {journal} {\bibinfo  {journal} {Phys. Rev. D}\ }\textbf {\bibinfo {volume}
  {83}},\ \bibinfo {pages} {034016} (\bibinfo {year} {2011})},\ \Eprint
  {https://arxiv.org/abs/1012.1291} {arXiv:1012.1291 [hep-ph]} \BibitemShut
  {NoStop}%
\bibitem [{\citenamefont {Backes}, \citenamefont {Marquezb},\ and\
  \citenamefont {Menezes}(2021)}]{Backes-2021mdt}%
  \BibitemOpen
  \bibfield  {author} {\bibinfo {author} {\bibfnamefont {B.~C.~T.}\
  \bibnamefont {Backes}}, \bibinfo {author} {\bibfnamefont {K.~D.}\
  \bibnamefont {Marquezb}},\ and\ \bibinfo {author} {\bibfnamefont {D.~P.}\
  \bibnamefont {Menezes}},\ }\href
  {https://doi.org/10.1140/epja/s10050-021-00544-2} {\bibfield  {journal}
  {\bibinfo  {journal} {Eur. Phys. J. A}\ }\textbf {\bibinfo {volume} {57}},\
  \bibinfo {pages} {229} (\bibinfo {year} {2021})},\ \Eprint
  {https://arxiv.org/abs/2103.14733} {arXiv:2103.14733 [nucl-th]} \BibitemShut
  {NoStop}%
\bibitem [{\citenamefont {Peshier}\ \emph {et~al.}(1994)\citenamefont
  {Peshier}, \citenamefont {Kampfer}, \citenamefont {Pavlenko},\ and\
  \citenamefont {Soff}}]{Peshier-1994zf}%
  \BibitemOpen
  \bibfield  {author} {\bibinfo {author} {\bibfnamefont {A.}~\bibnamefont
  {Peshier}}, \bibinfo {author} {\bibfnamefont {B.}~\bibnamefont {Kampfer}},
  \bibinfo {author} {\bibfnamefont {O.~P.}\ \bibnamefont {Pavlenko}},\ and\
  \bibinfo {author} {\bibfnamefont {G.}~\bibnamefont {Soff}},\ }\href
  {https://doi.org/10.1016/0370-2693(94)90969-5} {\bibfield  {journal}
  {\bibinfo  {journal} {Phys. Lett. B}\ }\textbf {\bibinfo {volume} {337}},\
  \bibinfo {pages} {235} (\bibinfo {year} {1994})}\BibitemShut {NoStop}%
\bibitem [{\citenamefont {Gorenstein}\ and\ \citenamefont
  {Yang}(1995)}]{Gorenstein-1995vm}%
  \BibitemOpen
  \bibfield  {author} {\bibinfo {author} {\bibfnamefont {M.~I.}\ \bibnamefont
  {Gorenstein}}\ and\ \bibinfo {author} {\bibfnamefont {S.-N.}\ \bibnamefont
  {Yang}},\ }\href {https://doi.org/10.1103/PhysRevD.52.5206} {\bibfield
  {journal} {\bibinfo  {journal} {Phys. Rev. D}\ }\textbf {\bibinfo {volume}
  {52}},\ \bibinfo {pages} {5206} (\bibinfo {year} {1995})}\BibitemShut
  {NoStop}%
\bibitem [{\citenamefont {Bannur}(2007)}]{Bannur-2006js}%
  \BibitemOpen
  \bibfield  {author} {\bibinfo {author} {\bibfnamefont {V.~M.}\ \bibnamefont
  {Bannur}},\ }\href {https://doi.org/10.1103/PhysRevC.75.044905} {\bibfield
  {journal} {\bibinfo  {journal} {Phys. Rev. C}\ }\textbf {\bibinfo {volume}
  {75}},\ \bibinfo {pages} {044905} (\bibinfo {year} {2007})},\ \Eprint
  {https://arxiv.org/abs/hep-ph/0609188} {arXiv:hep-ph/0609188} \BibitemShut
  {NoStop}%
\bibitem [{\citenamefont {Alba}\ \emph {et~al.}(2014)\citenamefont {Alba},
  \citenamefont {Alberico}, \citenamefont {Bluhm}, \citenamefont {Greco},
  \citenamefont {Ratti},\ and\ \citenamefont {Ruggieri}}]{Alba-2014lda}%
  \BibitemOpen
  \bibfield  {author} {\bibinfo {author} {\bibfnamefont {P.}~\bibnamefont
  {Alba}}, \bibinfo {author} {\bibfnamefont {W.}~\bibnamefont {Alberico}},
  \bibinfo {author} {\bibfnamefont {M.}~\bibnamefont {Bluhm}}, \bibinfo
  {author} {\bibfnamefont {V.}~\bibnamefont {Greco}}, \bibinfo {author}
  {\bibfnamefont {C.}~\bibnamefont {Ratti}},\ and\ \bibinfo {author}
  {\bibfnamefont {M.}~\bibnamefont {Ruggieri}},\ }\href
  {https://doi.org/10.1016/j.nuclphysa.2014.11.011} {\bibfield  {journal}
  {\bibinfo  {journal} {Nucl. Phys. A}\ }\textbf {\bibinfo {volume} {934}},\
  \bibinfo {pages} {41} (\bibinfo {year} {2014})},\ \Eprint
  {https://arxiv.org/abs/1402.6213} {arXiv:1402.6213 [hep-ph]} \BibitemShut
  {NoStop}%
\bibitem [{\citenamefont {Xu}\ \emph {et~al.}(2015{\natexlab{a}})\citenamefont
  {Xu}, \citenamefont {Peng}, \citenamefont {Lu},\ and\ \citenamefont
  {Cui}}]{Xu-2014zea}%
  \BibitemOpen
  \bibfield  {author} {\bibinfo {author} {\bibfnamefont {J.-F.}\ \bibnamefont
  {Xu}}, \bibinfo {author} {\bibfnamefont {G.-X.}\ \bibnamefont {Peng}},
  \bibinfo {author} {\bibfnamefont {Z.-Y.}\ \bibnamefont {Lu}},\ and\ \bibinfo
  {author} {\bibfnamefont {S.-S.}\ \bibnamefont {Cui}},\ }\href
  {https://doi.org/10.1007/s11433-014-5599-6} {\bibfield  {journal} {\bibinfo
  {journal} {Sci. China Phys. Mech. Astron.}\ }\textbf {\bibinfo {volume}
  {58}},\ \bibinfo {pages} {042001} (\bibinfo {year}
  {2015}{\natexlab{a}})}\BibitemShut {NoStop}%
\bibitem [{\citenamefont {Xu}\ \emph {et~al.}(2015{\natexlab{b}})\citenamefont
  {Xu}, \citenamefont {Peng}, \citenamefont {Liu}, \citenamefont {Hou},\ and\
  \citenamefont {Chen}}]{Xu-2015wya}%
  \BibitemOpen
  \bibfield  {author} {\bibinfo {author} {\bibfnamefont {J.~F.}\ \bibnamefont
  {Xu}}, \bibinfo {author} {\bibfnamefont {G.~X.}\ \bibnamefont {Peng}},
  \bibinfo {author} {\bibfnamefont {F.}~\bibnamefont {Liu}}, \bibinfo {author}
  {\bibfnamefont {D.-F.}\ \bibnamefont {Hou}},\ and\ \bibinfo {author}
  {\bibfnamefont {L.-W.}\ \bibnamefont {Chen}},\ }\href
  {https://doi.org/10.1103/PhysRevD.92.025025} {\bibfield  {journal} {\bibinfo
  {journal} {Phys. Rev. D}\ }\textbf {\bibinfo {volume} {92}},\ \bibinfo
  {pages} {025025} (\bibinfo {year} {2015}{\natexlab{b}})},\ \Eprint
  {https://arxiv.org/abs/1512.08229} {arXiv:1512.08229 [hep-ph]} \BibitemShut
  {NoStop}%
\bibitem [{\citenamefont {Xia}\ \emph {et~al.}(2014)\citenamefont {Xia},
  \citenamefont {Peng}, \citenamefont {Chen}, \citenamefont {Lu},\ and\
  \citenamefont {Xu}}]{Xia-2014zaa}%
  \BibitemOpen
  \bibfield  {author} {\bibinfo {author} {\bibfnamefont {C.~J.}\ \bibnamefont
  {Xia}}, \bibinfo {author} {\bibfnamefont {G.~X.}\ \bibnamefont {Peng}},
  \bibinfo {author} {\bibfnamefont {S.~W.}\ \bibnamefont {Chen}}, \bibinfo
  {author} {\bibfnamefont {Z.~Y.}\ \bibnamefont {Lu}},\ and\ \bibinfo {author}
  {\bibfnamefont {J.~F.}\ \bibnamefont {Xu}},\ }\href
  {https://doi.org/10.1103/PhysRevD.89.105027} {\bibfield  {journal} {\bibinfo
  {journal} {Phys. Rev. D}\ }\textbf {\bibinfo {volume} {89}},\ \bibinfo
  {pages} {105027} (\bibinfo {year} {2014})},\ \Eprint
  {https://arxiv.org/abs/1405.3037} {arXiv:1405.3037 [hep-ph]} \BibitemShut
  {NoStop}%
\bibitem [{\citenamefont {Zhu}\ \emph {et~al.}(2019)\citenamefont {Zhu},
  \citenamefont {Li}, \citenamefont {Hu},\ and\ \citenamefont
  {Shen}}]{Zhu-2018vwn}%
  \BibitemOpen
  \bibfield  {author} {\bibinfo {author} {\bibfnamefont {Z.-Y.}\ \bibnamefont
  {Zhu}}, \bibinfo {author} {\bibfnamefont {A.}~\bibnamefont {Li}}, \bibinfo
  {author} {\bibfnamefont {J.-N.}\ \bibnamefont {Hu}},\ and\ \bibinfo {author}
  {\bibfnamefont {H.}~\bibnamefont {Shen}},\ }\href
  {https://doi.org/10.1103/PhysRevC.99.025804} {\bibfield  {journal} {\bibinfo
  {journal} {Phys. Rev. C}\ }\textbf {\bibinfo {volume} {99}},\ \bibinfo
  {pages} {025804} (\bibinfo {year} {2019})},\ \Eprint
  {https://arxiv.org/abs/1805.04678} {arXiv:1805.04678 [nucl-th]} \BibitemShut
  {NoStop}%
\bibitem [{\citenamefont {Peshier}, \citenamefont {Kampfer},\ and\
  \citenamefont {Soff}(2000)}]{Peshier-1999ww}%
  \BibitemOpen
  \bibfield  {author} {\bibinfo {author} {\bibfnamefont {A.}~\bibnamefont
  {Peshier}}, \bibinfo {author} {\bibfnamefont {B.}~\bibnamefont {Kampfer}},\
  and\ \bibinfo {author} {\bibfnamefont {G.}~\bibnamefont {Soff}},\ }\href
  {https://doi.org/10.1103/PhysRevC.61.045203} {\bibfield  {journal} {\bibinfo
  {journal} {Phys. Rev. C}\ }\textbf {\bibinfo {volume} {61}},\ \bibinfo
  {pages} {045203} (\bibinfo {year} {2000})},\ \Eprint
  {https://arxiv.org/abs/hep-ph/9911474} {arXiv:hep-ph/9911474 [hep-ph]}
  \BibitemShut {NoStop}%
\bibitem [{\citenamefont {Ma}\ \emph {et~al.}(2019)\citenamefont {Ma},
  \citenamefont {Lin}, \citenamefont {Qian}, \citenamefont {Hama},\ and\
  \citenamefont {Kodama}}]{Ma-2018bwf}%
  \BibitemOpen
  \bibfield  {author} {\bibinfo {author} {\bibfnamefont {H.-H.}\ \bibnamefont
  {Ma}}, \bibinfo {author} {\bibfnamefont {K.}~\bibnamefont {Lin}}, \bibinfo
  {author} {\bibfnamefont {W.-L.}\ \bibnamefont {Qian}}, \bibinfo {author}
  {\bibfnamefont {Y.}~\bibnamefont {Hama}},\ and\ \bibinfo {author}
  {\bibfnamefont {T.}~\bibnamefont {Kodama}},\ }\href
  {https://doi.org/10.1103/PhysRevC.100.015206} {\bibfield  {journal} {\bibinfo
   {journal} {Phys. Rev. C}\ }\textbf {\bibinfo {volume} {100}},\ \bibinfo
  {pages} {015206} (\bibinfo {year} {2019})},\ \Eprint
  {https://arxiv.org/abs/1804.05376} {arXiv:1804.05376 [hep-ph]} \BibitemShut
  {NoStop}%
\bibitem [{\citenamefont {Schertler}, \citenamefont {Greiner},\ and\
  \citenamefont {Thoma}(1997)}]{Schertler-1996tq}%
  \BibitemOpen
  \bibfield  {author} {\bibinfo {author} {\bibfnamefont {K.}~\bibnamefont
  {Schertler}}, \bibinfo {author} {\bibfnamefont {C.}~\bibnamefont {Greiner}},\
  and\ \bibinfo {author} {\bibfnamefont {M.~H.}\ \bibnamefont {Thoma}},\
  }\bibfield  {booktitle} {\emph {\bibinfo {booktitle} {{Proceedings, 4th
  International Conference on Radioactive Nuclear Beams (RNB 4): Ohmiya, Japan,
  June 3-7, 1996}}},\ }\href {https://doi.org/10.1016/S0375-9474(97)00014-6}
  {\bibfield  {journal} {\bibinfo  {journal} {Nucl. Phys. A}\ }\textbf
  {\bibinfo {volume} {616}},\ \bibinfo {pages} {659} (\bibinfo {year}
  {1997})},\ \Eprint {https://arxiv.org/abs/hep-ph/9611305}
  {arXiv:hep-ph/9611305 [hep-ph]} \BibitemShut {NoStop}%
\bibitem [{\citenamefont {Schertler}\ \emph {et~al.}(1998)\citenamefont
  {Schertler}, \citenamefont {Greiner}, \citenamefont {Sahu},\ and\
  \citenamefont {Thoma}}]{Schertler-1997za}%
  \BibitemOpen
  \bibfield  {author} {\bibinfo {author} {\bibfnamefont {K.}~\bibnamefont
  {Schertler}}, \bibinfo {author} {\bibfnamefont {C.}~\bibnamefont {Greiner}},
  \bibinfo {author} {\bibfnamefont {P.}~\bibnamefont {Sahu}},\ and\ \bibinfo
  {author} {\bibfnamefont {M.}~\bibnamefont {Thoma}},\ }\href
  {https://doi.org/10.1016/S0375-9474(98)00330-3} {\bibfield  {journal}
  {\bibinfo  {journal} {Nucl. Phys. A}\ }\textbf {\bibinfo {volume} {637}},\
  \bibinfo {pages} {451} (\bibinfo {year} {1998})},\ \Eprint
  {https://arxiv.org/abs/astro-ph/9712165} {arXiv:astro-ph/9712165}
  \BibitemShut {NoStop}%
\bibitem [{\citenamefont {Wen}\ \emph {et~al.}(2009)\citenamefont {Wen},
  \citenamefont {Feng}, \citenamefont {Li},\ and\ \citenamefont
  {Peng}}]{Wen-2009zza}%
  \BibitemOpen
  \bibfield  {author} {\bibinfo {author} {\bibfnamefont {X.~J.}\ \bibnamefont
  {Wen}}, \bibinfo {author} {\bibfnamefont {Z.~Q.}\ \bibnamefont {Feng}},
  \bibinfo {author} {\bibfnamefont {N.}~\bibnamefont {Li}},\ and\ \bibinfo
  {author} {\bibfnamefont {G.~X.}\ \bibnamefont {Peng}},\ }\href
  {https://doi.org/10.1088/0954-3899/36/2/025011} {\bibfield  {journal}
  {\bibinfo  {journal} {J. Phys. G}\ }\textbf {\bibinfo {volume} {36}},\
  \bibinfo {pages} {025011} (\bibinfo {year} {2009})}\BibitemShut {NoStop}%
\bibitem [{\citenamefont {Wen}\ \emph {et~al.}(2010)\citenamefont {Wen},
  \citenamefont {Li}, \citenamefont {Liang},\ and\ \citenamefont
  {Peng}}]{Wen-2010zz}%
  \BibitemOpen
  \bibfield  {author} {\bibinfo {author} {\bibfnamefont {X.~J.}\ \bibnamefont
  {Wen}}, \bibinfo {author} {\bibfnamefont {J.~Y.}\ \bibnamefont {Li}},
  \bibinfo {author} {\bibfnamefont {J.~Q.}\ \bibnamefont {Liang}},\ and\
  \bibinfo {author} {\bibfnamefont {G.~X.}\ \bibnamefont {Peng}},\ }\href
  {https://doi.org/10.1103/PhysRevC.82.025809} {\bibfield  {journal} {\bibinfo
  {journal} {Phys. Rev.C}\ }\textbf {\bibinfo {volume} {82}},\ \bibinfo {pages}
  {025809} (\bibinfo {year} {2010})}\BibitemShut {NoStop}%
\bibitem [{\citenamefont {Wen}, \citenamefont {Yang},\ and\ \citenamefont
  {Su}(2011)}]{Wen-2011zz}%
  \BibitemOpen
  \bibfield  {author} {\bibinfo {author} {\bibfnamefont {X.-J.}\ \bibnamefont
  {Wen}}, \bibinfo {author} {\bibfnamefont {D.-H.}\ \bibnamefont {Yang}},\ and\
  \bibinfo {author} {\bibfnamefont {S.-Z.}\ \bibnamefont {Su}},\ }\href
  {https://doi.org/10.1088/0954-3899/38/11/115001} {\bibfield  {journal}
  {\bibinfo  {journal} {J. Phys. G}\ }\textbf {\bibinfo {volume} {38}},\
  \bibinfo {pages} {115001} (\bibinfo {year} {2011})}\BibitemShut {NoStop}%
\bibitem [{\citenamefont {Lu}\ \emph {et~al.}(2016{\natexlab{a}})\citenamefont
  {Lu}, \citenamefont {Peng}, \citenamefont {Xu},\ and\ \citenamefont
  {Zhang}}]{Lu-2016fki}%
  \BibitemOpen
  \bibfield  {author} {\bibinfo {author} {\bibfnamefont {Z.-Y.}\ \bibnamefont
  {Lu}}, \bibinfo {author} {\bibfnamefont {G.-X.}\ \bibnamefont {Peng}},
  \bibinfo {author} {\bibfnamefont {J.-F.}\ \bibnamefont {Xu}},\ and\ \bibinfo
  {author} {\bibfnamefont {S.-P.}\ \bibnamefont {Zhang}},\ }\href
  {https://doi.org/10.1007/s11433-015-0524-2} {\bibfield  {journal} {\bibinfo
  {journal} {Sci. China Phys. Mech. Astron.}\ }\textbf {\bibinfo {volume}
  {59}},\ \bibinfo {pages} {662001} (\bibinfo {year}
  {2016}{\natexlab{a}})}\BibitemShut {NoStop}%
\bibitem [{\citenamefont {Chakrabarty}(1991)}]{Chakrabarty-1991ui}%
  \BibitemOpen
  \bibfield  {author} {\bibinfo {author} {\bibfnamefont {S.}~\bibnamefont
  {Chakrabarty}},\ }\href {https://doi.org/10.1103/PhysRevD.43.627} {\bibfield
  {journal} {\bibinfo  {journal} {Phys. Rev. D}\ }\textbf {\bibinfo {volume}
  {43}},\ \bibinfo {pages} {627} (\bibinfo {year} {1991})}\BibitemShut
  {NoStop}%
\bibitem [{\citenamefont {Peng}\ \emph {et~al.}(2000)\citenamefont {Peng},
  \citenamefont {Chiang}, \citenamefont {Yang}, \citenamefont {Li},\ and\
  \citenamefont {Liu}}]{Peng-1999gh}%
  \BibitemOpen
  \bibfield  {author} {\bibinfo {author} {\bibfnamefont {G.~X.}\ \bibnamefont
  {Peng}}, \bibinfo {author} {\bibfnamefont {H.~C.}\ \bibnamefont {Chiang}},
  \bibinfo {author} {\bibfnamefont {J.~J.}\ \bibnamefont {Yang}}, \bibinfo
  {author} {\bibfnamefont {L.}~\bibnamefont {Li}},\ and\ \bibinfo {author}
  {\bibfnamefont {B.}~\bibnamefont {Liu}},\ }\href
  {https://doi.org/10.1103/PhysRevC.61.015201} {\bibfield  {journal} {\bibinfo
  {journal} {Phys. Rev. C}\ }\textbf {\bibinfo {volume} {61}},\ \bibinfo
  {pages} {015201} (\bibinfo {year} {2000})},\ \Eprint
  {https://arxiv.org/abs/hep-ph/9911222} {arXiv:hep-ph/9911222 [hep-ph]}
  \BibitemShut {NoStop}%
\bibitem [{\citenamefont {Wen}\ \emph {et~al.}(2005)\citenamefont {Wen},
  \citenamefont {Zhong}, \citenamefont {Peng}, \citenamefont {Shen},\ and\
  \citenamefont {Ning}}]{Wen-2005uf}%
  \BibitemOpen
  \bibfield  {author} {\bibinfo {author} {\bibfnamefont {X.~J.}\ \bibnamefont
  {Wen}}, \bibinfo {author} {\bibfnamefont {X.~H.}\ \bibnamefont {Zhong}},
  \bibinfo {author} {\bibfnamefont {G.~X.}\ \bibnamefont {Peng}}, \bibinfo
  {author} {\bibfnamefont {P.~N.}\ \bibnamefont {Shen}},\ and\ \bibinfo
  {author} {\bibfnamefont {P.~Z.}\ \bibnamefont {Ning}},\ }\href
  {https://doi.org/10.1103/PhysRevC.72.015204} {\bibfield  {journal} {\bibinfo
  {journal} {Phys. Rev. C}\ }\textbf {\bibinfo {volume} {72}},\ \bibinfo
  {pages} {015204} (\bibinfo {year} {2005})},\ \Eprint
  {https://arxiv.org/abs/hep-ph/0506050} {arXiv:hep-ph/0506050 [hep-ph]}
  \BibitemShut {NoStop}%
\bibitem [{\citenamefont {Lu}\ \emph {et~al.}(2016{\natexlab{b}})\citenamefont
  {Lu}, \citenamefont {Peng}, \citenamefont {Zhang}, \citenamefont {Ruggieri},\
  and\ \citenamefont {Greco}}]{Lu-2016jsv}%
  \BibitemOpen
  \bibfield  {author} {\bibinfo {author} {\bibfnamefont {Z.-Y.}\ \bibnamefont
  {Lu}}, \bibinfo {author} {\bibfnamefont {G.-X.}\ \bibnamefont {Peng}},
  \bibinfo {author} {\bibfnamefont {S.-P.}\ \bibnamefont {Zhang}}, \bibinfo
  {author} {\bibfnamefont {M.}~\bibnamefont {Ruggieri}},\ and\ \bibinfo
  {author} {\bibfnamefont {V.}~\bibnamefont {Greco}},\ }\href
  {https://doi.org/10.1007/s41365-016-0148-9} {\bibfield  {journal} {\bibinfo
  {journal} {Nucl. Sci. Tech.}\ }\textbf {\bibinfo {volume} {27}},\ \bibinfo
  {pages} {148} (\bibinfo {year} {2016}{\natexlab{b}})}\BibitemShut {NoStop}%
\bibitem [{\citenamefont {Chu}\ and\ \citenamefont {Chen}(2014)}]{Chu-2012rd}%
  \BibitemOpen
  \bibfield  {author} {\bibinfo {author} {\bibfnamefont {P.-C.}\ \bibnamefont
  {Chu}}\ and\ \bibinfo {author} {\bibfnamefont {L.-W.}\ \bibnamefont {Chen}},\
  }\href {https://doi.org/10.1088/0004-637X/780/2/135} {\bibfield  {journal}
  {\bibinfo  {journal} {Astrophys. J.}\ }\textbf {\bibinfo {volume} {780}},\
  \bibinfo {pages} {135} (\bibinfo {year} {2014})},\ \Eprint
  {https://arxiv.org/abs/1212.1388} {arXiv:1212.1388 [astro-ph.SR]}
  \BibitemShut {NoStop}%
\bibitem [{\citenamefont {Vija}\ and\ \citenamefont
  {Thoma}(1995)}]{Vija-1994is}%
  \BibitemOpen
  \bibfield  {author} {\bibinfo {author} {\bibfnamefont {H.}~\bibnamefont
  {Vija}}\ and\ \bibinfo {author} {\bibfnamefont {M.~H.}\ \bibnamefont
  {Thoma}},\ }\href {https://doi.org/10.1016/0370-2693(94)01378-P} {\bibfield
  {journal} {\bibinfo  {journal} {Phys. Lett. B}\ }\textbf {\bibinfo {volume}
  {342}},\ \bibinfo {pages} {212} (\bibinfo {year} {1995})},\ \Eprint
  {https://arxiv.org/abs/hep-ph/9409246} {arXiv:hep-ph/9409246 [hep-ph]}
  \BibitemShut {NoStop}%
\bibitem [{\citenamefont {Weldon}(1982)}]{Weldon-1982aq}%
  \BibitemOpen
  \bibfield  {author} {\bibinfo {author} {\bibfnamefont {H.}~\bibnamefont
  {Weldon}},\ }\href {https://doi.org/10.1103/PhysRevD.26.1394} {\bibfield
  {journal} {\bibinfo  {journal} {Phys. Rev. D}\ }\textbf {\bibinfo {volume}
  {26}},\ \bibinfo {pages} {1394} (\bibinfo {year} {1982})}\BibitemShut
  {NoStop}%
\bibitem [{\citenamefont {Zhang}\ \emph {et~al.}(2021)\citenamefont {Zhang},
  \citenamefont {Chu}, \citenamefont {Li}, \citenamefont {Liu},\ and\
  \citenamefont {Zhang}}]{Zhang-2021qhl}%
  \BibitemOpen
  \bibfield  {author} {\bibinfo {author} {\bibfnamefont {Z.}~\bibnamefont
  {Zhang}}, \bibinfo {author} {\bibfnamefont {P.-C.}\ \bibnamefont {Chu}},
  \bibinfo {author} {\bibfnamefont {X.-H.}\ \bibnamefont {Li}}, \bibinfo
  {author} {\bibfnamefont {H.}~\bibnamefont {Liu}},\ and\ \bibinfo {author}
  {\bibfnamefont {X.-M.}\ \bibnamefont {Zhang}},\ }\href
  {https://doi.org/10.1103/PhysRevD.103.103021} {\bibfield  {journal} {\bibinfo
   {journal} {Phys. Rev. D}\ }\textbf {\bibinfo {volume} {103}},\ \bibinfo
  {pages} {103021} (\bibinfo {year} {2021})}\BibitemShut {NoStop}%
\bibitem [{\citenamefont {Patra}\ and\ \citenamefont
  {Singh}(1996)}]{Patra-1995gs}%
  \BibitemOpen
  \bibfield  {author} {\bibinfo {author} {\bibfnamefont {B.~K.}\ \bibnamefont
  {Patra}}\ and\ \bibinfo {author} {\bibfnamefont {C.~P.}\ \bibnamefont
  {Singh}},\ }\href {https://doi.org/10.1103/PhysRevD.54.3551} {\bibfield
  {journal} {\bibinfo  {journal} {Phys. Rev. D}\ }\textbf {\bibinfo {volume}
  {54}},\ \bibinfo {pages} {3551} (\bibinfo {year} {1996})}\BibitemShut
  {NoStop}%
\bibitem [{\citenamefont {Broderick}, \citenamefont {Prakash},\ and\
  \citenamefont {Lattimer}(2000)}]{Broderick-2000pe}%
  \BibitemOpen
  \bibfield  {author} {\bibinfo {author} {\bibfnamefont {A.}~\bibnamefont
  {Broderick}}, \bibinfo {author} {\bibfnamefont {M.}~\bibnamefont {Prakash}},\
  and\ \bibinfo {author} {\bibfnamefont {J.}~\bibnamefont {Lattimer}},\ }\href
  {https://doi.org/10.1086/309010} {\bibfield  {journal} {\bibinfo  {journal}
  {Astrophys. J.}\ }\textbf {\bibinfo {volume} {537}},\ \bibinfo {pages} {351}
  (\bibinfo {year} {2000})},\ \Eprint {https://arxiv.org/abs/astro-ph/0001537}
  {arXiv:astro-ph/0001537} \BibitemShut {NoStop}%
\bibitem [{\citenamefont {Strickland}, \citenamefont {Dexheimer},\ and\
  \citenamefont {Menezes}(2012)}]{Strickland-2012vu}%
  \BibitemOpen
  \bibfield  {author} {\bibinfo {author} {\bibfnamefont {M.}~\bibnamefont
  {Strickland}}, \bibinfo {author} {\bibfnamefont {V.}~\bibnamefont
  {Dexheimer}},\ and\ \bibinfo {author} {\bibfnamefont {D.~P.}\ \bibnamefont
  {Menezes}},\ }\href {https://doi.org/10.1103/PhysRevD.86.125032} {\bibfield
  {journal} {\bibinfo  {journal} {Phys. Rev. D}\ }\textbf {\bibinfo {volume}
  {86}},\ \bibinfo {pages} {125032} (\bibinfo {year} {2012})},\ \Eprint
  {https://arxiv.org/abs/1209.3276} {arXiv:1209.3276 [nucl-th]} \BibitemShut
  {NoStop}%
\bibitem [{\citenamefont {Wen}\ and\ \citenamefont
  {Liang}(2016)}]{Wen-2016atg}%
  \BibitemOpen
  \bibfield  {author} {\bibinfo {author} {\bibfnamefont {X.-J.}\ \bibnamefont
  {Wen}}\ and\ \bibinfo {author} {\bibfnamefont {J.-J.}\ \bibnamefont
  {Liang}},\ }\href {https://doi.org/10.1103/PhysRevD.94.014005} {\bibfield
  {journal} {\bibinfo  {journal} {Phys. Rev. D}\ }\textbf {\bibinfo {volume}
  {94}},\ \bibinfo {pages} {014005} (\bibinfo {year} {2016})},\ \Eprint
  {https://arxiv.org/abs/1606.09336} {arXiv:1606.09336 [hep-ph]} \BibitemShut
  {NoStop}%
\bibitem [{\citenamefont {Perez-Garcia}, \citenamefont {Providencia},\ and\
  \citenamefont {Rabhi}(2011)}]{Perez-Garcia-2011czx}%
  \BibitemOpen
  \bibfield  {author} {\bibinfo {author} {\bibfnamefont {M.~A.}\ \bibnamefont
  {Perez-Garcia}}, \bibinfo {author} {\bibfnamefont {C.}~\bibnamefont
  {Providencia}},\ and\ \bibinfo {author} {\bibfnamefont {A.}~\bibnamefont
  {Rabhi}},\ }\href {https://doi.org/10.1103/PhysRevC.84.045803} {\bibfield
  {journal} {\bibinfo  {journal} {Phys. Rev. C}\ }\textbf {\bibinfo {volume}
  {84}},\ \bibinfo {pages} {045803} (\bibinfo {year} {2011})},\ \Eprint
  {https://arxiv.org/abs/1101.1656} {arXiv:1101.1656 [nucl-th]} \BibitemShut
  {NoStop}%
\bibitem [{\citenamefont {Avancini}, \citenamefont {Menezes},\ and\
  \citenamefont {Providencia}(2011)}]{Avancini-2011zz}%
  \BibitemOpen
  \bibfield  {author} {\bibinfo {author} {\bibfnamefont {S.~S.}\ \bibnamefont
  {Avancini}}, \bibinfo {author} {\bibfnamefont {D.~P.}\ \bibnamefont
  {Menezes}},\ and\ \bibinfo {author} {\bibfnamefont {C.}~\bibnamefont
  {Providencia}},\ }\href {https://doi.org/10.1103/PhysRevC.83.065805}
  {\bibfield  {journal} {\bibinfo  {journal} {Phys. Rev. C}\ }\textbf {\bibinfo
  {volume} {83}},\ \bibinfo {pages} {065805} (\bibinfo {year}
  {2011})}\BibitemShut {NoStop}%
\bibitem [{\citenamefont {Rabhi}\ \emph {et~al.}(2015)\citenamefont {Rabhi},
  \citenamefont {P\'erez-Garc\'\i{}a}, \citenamefont {Provid\^encia},\ and\
  \citenamefont {Vida\~na}}]{Rabhi-2014sza}%
  \BibitemOpen
  \bibfield  {author} {\bibinfo {author} {\bibfnamefont {A.}~\bibnamefont
  {Rabhi}}, \bibinfo {author} {\bibfnamefont {M.~A.}\ \bibnamefont
  {P\'erez-Garc\'\i{}a}}, \bibinfo {author} {\bibfnamefont {C.}~\bibnamefont
  {Provid\^encia}},\ and\ \bibinfo {author} {\bibfnamefont {I.}~\bibnamefont
  {Vida\~na}},\ }\href {https://doi.org/10.1103/PhysRevC.91.045803} {\bibfield
  {journal} {\bibinfo  {journal} {Phys. Rev. C}\ }\textbf {\bibinfo {volume}
  {91}},\ \bibinfo {pages} {045803} (\bibinfo {year} {2015})},\ \Eprint
  {https://arxiv.org/abs/1410.2748} {arXiv:1410.2748 [nucl-th]} \BibitemShut
  {NoStop}%
\bibitem [{\citenamefont {Felipe}\ \emph {et~al.}(2008)\citenamefont {Felipe},
  \citenamefont {Martinez}, \citenamefont {Rojas},\ and\ \citenamefont
  {Orsaria}}]{Felipe-2007vb}%
  \BibitemOpen
  \bibfield  {author} {\bibinfo {author} {\bibfnamefont {R.~G.}\ \bibnamefont
  {Felipe}}, \bibinfo {author} {\bibfnamefont {A.~P.}\ \bibnamefont
  {Martinez}}, \bibinfo {author} {\bibfnamefont {H.~P.}\ \bibnamefont
  {Rojas}},\ and\ \bibinfo {author} {\bibfnamefont {M.}~\bibnamefont
  {Orsaria}},\ }\href {https://doi.org/10.1103/PhysRevC.77.015807} {\bibfield
  {journal} {\bibinfo  {journal} {Phys. Rev. C}\ }\textbf {\bibinfo {volume}
  {77}},\ \bibinfo {pages} {015807} (\bibinfo {year} {2008})},\ \Eprint
  {https://arxiv.org/abs/0709.1224} {arXiv:0709.1224 [astro-ph]} \BibitemShut
  {NoStop}%
\bibitem [{\citenamefont {Felipe}\ and\ \citenamefont
  {Martinez}(2009)}]{Felipe-2008cm}%
  \BibitemOpen
  \bibfield  {author} {\bibinfo {author} {\bibfnamefont {R.~G.}\ \bibnamefont
  {Felipe}}\ and\ \bibinfo {author} {\bibfnamefont {A.~P.}\ \bibnamefont
  {Martinez}},\ }\href {https://doi.org/10.1088/0954-3899/36/7/075202}
  {\bibfield  {journal} {\bibinfo  {journal} {J. Phys. G}\ }\textbf {\bibinfo
  {volume} {36}},\ \bibinfo {pages} {075202} (\bibinfo {year} {2009})},\
  \Eprint {https://arxiv.org/abs/0812.0337} {arXiv:0812.0337 [astro-ph]}
  \BibitemShut {NoStop}%
\bibitem [{\citenamefont {Isayev}\ and\ \citenamefont
  {Yang}(2013)}]{Isayev-2012sv}%
  \BibitemOpen
  \bibfield  {author} {\bibinfo {author} {\bibfnamefont {A.~A.}\ \bibnamefont
  {Isayev}}\ and\ \bibinfo {author} {\bibfnamefont {J.}~\bibnamefont {Yang}},\
  }\href {https://doi.org/10.1088/0954-3899/40/3/035105} {\bibfield  {journal}
  {\bibinfo  {journal} {J. Phys. G}\ }\textbf {\bibinfo {volume} {40}},\
  \bibinfo {pages} {035105} (\bibinfo {year} {2013})},\ \Eprint
  {https://arxiv.org/abs/1210.3322} {arXiv:1210.3322 [hep-ph]} \BibitemShut
  {NoStop}%
\bibitem [{\citenamefont {Alsing}, \citenamefont {Silva},\ and\ \citenamefont
  {Berti}(2018)}]{Alsing-2017bbc}%
  \BibitemOpen
  \bibfield  {author} {\bibinfo {author} {\bibfnamefont {J.}~\bibnamefont
  {Alsing}}, \bibinfo {author} {\bibfnamefont {H.~O.}\ \bibnamefont {Silva}},\
  and\ \bibinfo {author} {\bibfnamefont {E.}~\bibnamefont {Berti}},\ }\href
  {https://doi.org/10.1093/mnras/sty1065} {\bibfield  {journal} {\bibinfo
  {journal} {Mon. Not. Roy. Astron. Soc.}\ }\textbf {\bibinfo {volume} {478}},\
  \bibinfo {pages} {1377} (\bibinfo {year} {2018})},\ \Eprint
  {https://arxiv.org/abs/1709.07889} {arXiv:1709.07889 [astro-ph.HE]}
  \BibitemShut {NoStop}%
\bibitem [{\citenamefont {Tews}, \citenamefont {Margueron},\ and\ \citenamefont
  {Reddy}(2018)}]{Tews-2018iwm}%
  \BibitemOpen
  \bibfield  {author} {\bibinfo {author} {\bibfnamefont {I.}~\bibnamefont
  {Tews}}, \bibinfo {author} {\bibfnamefont {J.}~\bibnamefont {Margueron}},\
  and\ \bibinfo {author} {\bibfnamefont {S.}~\bibnamefont {Reddy}},\ }\href
  {https://doi.org/10.1103/PhysRevC.98.045804} {\bibfield  {journal} {\bibinfo
  {journal} {Phys. Rev. C}\ }\textbf {\bibinfo {volume} {98}},\ \bibinfo
  {pages} {045804} (\bibinfo {year} {2018})},\ \Eprint
  {https://arxiv.org/abs/1804.02783} {arXiv:1804.02783 [nucl-th]} \BibitemShut
  {NoStop}%
\bibitem [{\citenamefont {Reed}\ and\ \citenamefont
  {Horowitz}(2020)}]{Reed-2019ezm}%
  \BibitemOpen
  \bibfield  {author} {\bibinfo {author} {\bibfnamefont {B.}~\bibnamefont
  {Reed}}\ and\ \bibinfo {author} {\bibfnamefont {C.}~\bibnamefont
  {Horowitz}},\ }\href {https://doi.org/10.1103/PhysRevC.101.045803} {\bibfield
   {journal} {\bibinfo  {journal} {Phys. Rev. C}\ }\textbf {\bibinfo {volume}
  {101}},\ \bibinfo {pages} {045803} (\bibinfo {year} {2020})},\ \Eprint
  {https://arxiv.org/abs/1910.05463} {arXiv:1910.05463 [astro-ph.HE]}
  \BibitemShut {NoStop}%
\bibitem [{\citenamefont {Xia}\ \emph {et~al.}(2021)\citenamefont {Xia},
  \citenamefont {Zhu}, \citenamefont {Zhou},\ and\ \citenamefont
  {Li}}]{Xia-2019xax}%
  \BibitemOpen
  \bibfield  {author} {\bibinfo {author} {\bibfnamefont {C.}~\bibnamefont
  {Xia}}, \bibinfo {author} {\bibfnamefont {Z.}~\bibnamefont {Zhu}}, \bibinfo
  {author} {\bibfnamefont {X.}~\bibnamefont {Zhou}},\ and\ \bibinfo {author}
  {\bibfnamefont {A.}~\bibnamefont {Li}},\ }\href
  {https://doi.org/10.1088/1674-1137/abea0d} {\bibfield  {journal} {\bibinfo
  {journal} {Chin. Phys. C}\ }\textbf {\bibinfo {volume} {45}},\ \bibinfo
  {pages} {055104} (\bibinfo {year} {2021})},\ \Eprint
  {https://arxiv.org/abs/1906.00826} {arXiv:1906.00826 [nucl-th]} \BibitemShut
  {NoStop}%
\bibitem [{\citenamefont {Li}\ \emph {et~al.}(2021)\citenamefont {Li},
  \citenamefont {Miao}, \citenamefont {Han},\ and\ \citenamefont
  {Zhang}}]{Li-2021crp}%
  \BibitemOpen
  \bibfield  {author} {\bibinfo {author} {\bibfnamefont {A.}~\bibnamefont
  {Li}}, \bibinfo {author} {\bibfnamefont {Z.}~\bibnamefont {Miao}}, \bibinfo
  {author} {\bibfnamefont {S.}~\bibnamefont {Han}},\ and\ \bibinfo {author}
  {\bibfnamefont {B.}~\bibnamefont {Zhang}},\ }\href
  {https://doi.org/10.3847/1538-4357/abf355} {\bibfield  {journal} {\bibinfo
  {journal} {Astrophys. J.}\ }\textbf {\bibinfo {volume} {913}},\ \bibinfo
  {pages} {27} (\bibinfo {year} {2021})},\ \Eprint
  {https://arxiv.org/abs/2103.15119} {arXiv:2103.15119 [astro-ph.HE]}
  \BibitemShut {NoStop}%
\bibitem [{\citenamefont {Hippert}, \citenamefont {Fraga},\ and\ \citenamefont
  {Noronha}(2021)}]{Hippert-2021gfs}%
  \BibitemOpen
  \bibfield  {author} {\bibinfo {author} {\bibfnamefont {M.}~\bibnamefont
  {Hippert}}, \bibinfo {author} {\bibfnamefont {E.~S.}\ \bibnamefont {Fraga}},\
  and\ \bibinfo {author} {\bibfnamefont {J.}~\bibnamefont {Noronha}},\ }\href
  {https://doi.org/10.1103/PhysRevD.104.034011} {\bibfield  {journal} {\bibinfo
   {journal} {Phys. Rev. D}\ }\textbf {\bibinfo {volume} {104}},\ \bibinfo
  {pages} {034011} (\bibinfo {year} {2021})},\ \Eprint
  {https://arxiv.org/abs/2105.04535} {arXiv:2105.04535 [nucl-th]} \BibitemShut
  {NoStop}%
\end{thebibliography}%

\end{document}